\documentclass[12pt,preprint]{aastex}

\def\lessim{\mathrel{\hbox{\rlap{\hbox{\lower4pt\hbox{$\sim$}}}\hbox{$<$}}}}
\def\grtsim{\mathrel{\hbox{\rlap{\hbox{\lower4pt\hbox{$\sim$}}}\hbox{$>$}}}}

\shorttitle{Eclipses of V~Per}
\shortauthors{Shafter \& Misselt}

\begin{document}


\title{Modeling Eclipses in the Classical Nova V~Persei: The Role of
the Accretion Disk Rim}


\author{A. W. Shafter}
\affil{Astronomy Department, San Diego State University,
    San Diego, CA 92182}
\email{aws@nova.sdsu.edu}

\and

\author{K. A. Misselt}
\affil{Steward Observatory, Tucson, AZ 85719}
\email{misselt@as.arizona.edu}




\begin{abstract}
Multicolor ($BVRI$) light curves of the eclipsing classical nova
V~Per are presented, and
a total of twelve new eclipse timings are measured for the system.
When combined with previous eclipse timings from the literature,
these timings yield a revised ephemeris for the times of mid-eclipse
given by ${\rm JD}_{\odot} = 2,447,442.8260(1) + 0.107123474(3)~E$.
The eclipse profiles are analyzed with a parameter-fitting model
that assumes four sources
of luminosity: a white dwarf primary star, a main-sequence
secondary star, a flared accretion disk with a rim,
and a bright spot at the intersection
of the mass-transfer stream and the disk periphery. Model parameters include
the temperatures of the white dwarf ($T_1$) and the secondary star ($T_2$),
the radius ($R_{\mathrm{d}}$) and temperature ($T_{\mathrm{d}}$),
of the disk periphery, the inner disk radius ($R_{\mathrm{in}}$),
the disk power-law temperature exponent ($\alpha$) and thickness ($h_{\mathrm{r}}$),
and a bright spot temperature enhancement factor ($\chi_\mathrm{s}$).
A matrix of model solutions are computed, covering an extensive range of
plausible parameter values. The solution matrix is then explored to determine
the optimum values for the fitting parameters and their associated errors.
For models that treat the accretion disk as a flat structure without a
rim, optimum fits require that the disk have a flat temperature profile.
Although models with a truncated inner disk ($R_{\mathrm{in}}>>R_1$)
result in a steeper temperature profile, steady-state models with a
temperature profile characterized by $T(r)\propto~r^{-3/4}$
are found only for models with a significant disk rim.
A comparison of the observed brightness and color at mid-eclipse
with the photometric properties of the best-fitting model
suggests that V~Per lies at a distance of $\sim1$~kpc.

\end{abstract}





\section{Introduction}

V~Per is an 18th magnitude classical nova system (Nova Persei 1887),
which reached an estimated magnitude of
$m\sim4-5$ at the peak of its eruption (see McLaughlin 1946).
More than a century later, the system
was discovered to be eclipsing by Shafter \& Abbott (1989) who established
an orbital period of 2.57~hr. V~Per is of particular interest because
its orbital period places the system near the
middle of the 2--3~hr ``gap'' in the orbital period distribution of
cataclysmic variable stars (Kolb 1996; Ritter 1996).
In addition to the possibility that
V~Per could have been formed in the period gap, Shafter \& Abbott
discussed possible evolutionary scenarios that could have brought
V~Per to its current period under the assumption that the system evolved
from an initially longer orbital period.

Eclipsing cataclysmic variables are valuable
because they provide the opportunity to study the
the radiative properties of accretion disks through analysis of the
eclipse profiles. Generally, properties of the system are determined either
through parameter-fitting models or through maximum-entropy eclipse mapping
techniques (e.g. Horne 1985).
In the former case, properties of the accretion disk, such
as its radial temperature profile, must be specified in analytic form,
while the latter method allows for an arbitrary disk emissivity distribution,
and selects the ``simplest" (maximum entropy) disk emissivity map.

Shortly after V~Per was discovered to be eclipsing,
Wood et~al. (1992) analyzed the original monochromatic
photometric data from Shafter \& Abbott (1989) using the maximum entropy
eclipse-mapping technique. The principal result of their study was that
the accretion disk in V~Per had a relatively ``flat" radial temperature
profile that was considerably less steep than the
$T(r)\propto r^{-3/4}$ profile
predicted for a steady-state disk (e.g. Shakura \& Sunyaev 1973).
Despite the inconsistency with the expected steady-state relation,
the flat temperature profile seen in V~Per is not unusual.
Several eclipse mapping studies of novalike variables and
dwarf novae in eruption -- systems expected to harbor steady-state
disks -- have also exhibited surprisingly flat temperature
profiles (e.g. Rutten et al. 1992; Robinson et al. 1999; B\'{\i}r\'o 2000).
This discrepancy has been explained by Smak (1994) who
used an eclipse-mapping technique to analyze
model eclipse profiles from disks with and without rims, demonstrating that
the spurious temperature profiles found in previous studies of
high inclination systems
($\grtsim80^{\circ}$) were a result of a failure
to model the disk rim. More recently,
Knigge et al. (2000)
used synthetic eclipse profiles of a rimmed disk from
their study of DW~UMa to show that
self-occultation of the disk by its rim,
in addition to the contribution of the rim
to the total light, causes the eclipse profiles
to be more {\bf V}-shaped than in the case of a flat disk.
They argued following Smak (1994), that
if these light curves were then modeled assuming the disk is flat (as in
eclipse mapping), an erroneously flat temperature gradient would result.
In the case of V~Per, Wood et al. (1992) found that a model with a truncated
inner disk, such as what one might expect if 
the white dwarf primary were strongly magnetic, had a temperature
profile that was considerably steeper, similar to that expected for a
steady-state disk.

Table~1 summarizes the properties of the 15
cataclysmic variables with measured periods between 2.25~hr and 2.75~hr.
Two of the stars are classified as classical novae (one of which is V~Per),
eight as magnetic systems (i.e. DQ~Her or
AM~Her stars), two as novalike variables, two as SU~UMa dwarf
novae, and one as both a nova and a DQ~Her star.
Given that eight out of the 15 systems near the middle of the period gap,
are in fact magnetic (i.e. either DQ~Her or AM~Her systems), it would
not be surprising if Wood et~al. (1992) were correct, and
V~Per turned out to be a magnetic system as well.
In order to pursue this question further, and to explore
how the addition
of a disk rim would affect the model results, we have obtained multicolor
eclipse light curves of V~Per and analyzed the data with a new
parameter-fitting eclipse model. In this paper, we present the results
of that study.

\section{Observations}

Observations of V~Per were carried out in 2002 October and December using
the $2048^2$ CCD imager on the Steward Observatory Bok 2.3~m reflector.
A summary of observations
is presented in Table~2. To
increase the time sampling efficiency, only a small subsection of the
full array was read out.  The subsection was centered on V~Per and
included several nearby stars that were used as comparison objects for
differential photometry.  Typical time resolutions
were in the range of 40-60~sec.

The data were processed in a standard fashion (bias subtraction and
flat-fielding) using IRAF.\footnote{
IRAF (Image Reduction and Analysis Facility) is distributed by the
National Optical Astronomy Observatories, which are operated by AURA, Inc.,
under cooperative agreement with the National Science Foundation.}
The individual images were subsequently
aligned to a common coordinate system and aperture photometry was
performed through a 6$''$ aperture on V~Per and two nearby comparison
stars, one much brighter than V~Per and the other of approximately
equal brightness. Differential light curves of V~Per
were constructed with respect to the brighter comparison star and the
stability of the atmospheric conditions was monitored using the star
of comparable brightness.  In all cases, the monitor star showed no
long-term drift, and had a point-to-point scatter of less than 0.02
differential magnitudes.

To calibrate our observations, several standard stars were observed
on each night.  Zero points were determined on each night using mean
extinction values appropriate to the Kitt Peak site.  On several
nights, standards were observed over a sufficient range of airmass for
a full solution including the extinction coefficient in each
band. In all cases, we found extinction coefficients consistent with
the mean values.  All subsequently reported magnitudes were computed
using nightly determined zero points and the mean extinction. The
two comparison stars in the V~Per field were calibrated by
calibrating every individual frame of our observations on each night
and computing the mean. The comparison star used to calibrate the V~Per
light curves is located at RA=2:01:54.3, DEC=+56:44:38.7 (2000.0)
($\sim35''$ north of V~Per), and is characterized by
$B=16.90\pm0.03, V=16.11\pm0.03, R=15.60\pm0.03$, and $I=15.09\pm0.03$.
The twelve calibrated
eclipse light curves -- three each in $B$, $V$, $R$, and $I$ --
are shown in Figure~1. V~Per is characterized by an out-of-eclipse
magnitude of $V\simeq18.1\pm0.1$ and $B-V\simeq0.6\pm0.2$.
The luminosity of the
system appears to be relatively stable over the $\sim2$ month time span
covered by the observations, with the exception of the data obtained
on 29~Oct~2002, when V~Per appeared to be $\sim0.4$ mag fainter than usual.
Mean magnitudes and colors of V~Per, both in and out of eclipse, are
given in Table~3.

\section{Updated Ephemeris}

Eclipse timings for V~Per extend back more than 15 years to
the original eclipses discovered in the work of Shafter \& Abbott (1988).
Table~4 summarizes all available timings of mid-eclipse
from Shafter \& Abbott (1989), Dahm (1997), Katysheva et~al. (2002),
and from the present study.  Following earlier studies,
times of mid-eclipse for the twelve eclipses in the present study
have been determined
by fitting a parabola to the lower half of the eclipse data.
A linear least-squares fit of the mid-eclipse times from Table~4
yields the following ephemeris for V~Per:

\begin{equation}
T_{\mathrm{mid-eclipse}} = {\rm JD}_{\odot}~2,447,442.8260(1)+0.107123474(3)~E.
\end{equation}

Residuals of the individual eclipse timings with respect to this ephemeris
are also given in Table~4, and are plotted as a function of cycle number
in Figure~2. The period of V~Per appears to be remarkably stable, with
no evidence for any significant period change over the $\sim15$ years
that the eclipses of V~Per have been monitored.

\section{Eclipse Model}

The V~Per data have been analyzed using an improved version of the
eclipse modeling program
described in Shafter et~al. (2000, Paper~I), and
in Shafter \& Holland (2003, Paper~II).
The model considers four principal
sources of light from the system: the white dwarf,
the secondary star, the accretion disk, and the bright spot where the
inter-star mass transfer stream impacts the periphery of the disk.
The model light curve flux at a given orbital
phase is computed by summing the contribution from the secondary star
and the unocculted regions of the remaining three sources of radiation.
The fluxes from all four sources of radiation have been corrected
via the use of a linear limb-darkening law with coefficients, $\mu_{\lambda}$, given
in van Hamme (1993).
The program operates on the normalized light curve intensities,
so in effect, we fit only the shapes and depths of the eclipse profiles.
The program does, however, compute a model $B-V$ color from the assumed
blackbody fluxes using the transformations given in Matthews \& Sandage (1963),
which can then be compared with the observed $B-V$ color.

The major improvement in our model is in the treatment of the accretion disk,
which is no longer confined to the orbital plane.
The disk thickness perpendicular to the orbital plane
now varies linearly with radius, resulting in a rim at the disk periphery.
The flaring angle of the disk and the rim (half) height are parameterized
by a fitting parameter, $h_{\mathrm{r}} = h/R_{\mathrm{d}}$, where $h$ is the disk height
at its edge perpendicular to the orbital plane, and $R_{\mathrm{d}}$ is
the disk radius. As before, the disk is assumed to
radiate like an optically-thick blackbody, with
a temperature distribution
that follows a radial power-law profile given by:
\begin{equation}
T(r) = T_{\mathrm{d}} \Bigg({R_{\mathrm{d}} \over r}\Bigg)^{\alpha},
\end{equation}
where $T_{\mathrm{d}}$ is the temperature of the rim at the outer edge of the disk.
For an optically thick, steady-state disk,
a characteristic value of $\alpha=0.75$ is expected (Shakura \& Sunyaev 1973).

Another improvement to earlier versions of our model involves the treatment of
radiation from the secondary star.
In the current model,
the tidally-distorted shape of the Roche-lobe-filling secondary star
is fully taken into account, not only when
computing the occulted regions of the white dwarf, disk, and bright spot,
as in earlier versions of the code,
but also when computing the
contribution of the secondary star to the model flux.
The flux from the white dwarf component is computed
assuming it to be a spherical blackbody of temperature $T_1$.
The radii of the white dwarf and
secondary stars are computed from the orbital period and mass ratio
as described in Paper~I.

One of the significant advantages of the inclusion of a disk rim
is that the bright spot can be more easily modeled.
The presence of the disk rim provides both an isotropic and
an anisotropic component to the bright spot radiation,
so there is no need to parameterize these components separately as in
Paper~II. 
The isotropic component of the bright spot is produced as in Paper~I,
by radiation from
a region of the disk's surface defined by
the intersection of a circular area of radius $0.2~R_{\mathrm{d}}$
(centered on the point of intersection of the mass-transfer stream
and the disk perimeter) and the accretion disk. The anisotropic
component is produced by radiation from the corresponding
azimuthal region of the disk rim.
As before, the bright spot is assumed to radiate like a blackbody,
the bright spot temperature parameterized
by a multiplicative factor, $\chi_\mathrm{s}$, applied to the
local disk temperature in the bright spot region.

For a given mass ratio and orbital inclination,
the geometry of the eclipse is determined by dividing the primary's
Roche lobe into a matrix of cells having Cartesian coordinates ($x,y,z$),
and then computing an occultation kernel $O(x,y,x,\phi)$ for eclipse by the
lobe-filling secondary star. The $x$ and $y$ axes lie in the orbital plane
with the positive $x$ axis pointing toward the
secondary star, and the positive $y$ axis oriented $90^{\circ}$
counterclockwise as seen from the direction of the positive $z$ axis.
For a given
orbital phase, $\phi$, the occultation kernel takes on values of zero or
one, depending on whether or not the cell is eclipsed.
The monochromatic fluxes from the accretion disk at orbital
phase $\phi$ are then given by the following summations.
For the disk surface and disk rim, respectively, we have:
\begin{equation}
f^{\mathrm s}_{\lambda}(\phi)={\mathrm{cos}~i \over d^2} \sum_{x=-R_\mathrm{d}}^{x=R_\mathrm{d}} \sum_{y=-\sqrt{R_\mathrm{d}^2-x^2}}^{y=\sqrt{R_\mathrm{d}^2-x^2}} I^{\mathrm s}_{\lambda}(x,y)~\Delta x~\Delta y~O(x,y,z=h_\mathrm{r}\sqrt{x^2+y^2},\phi),
\end{equation}
\begin{equation}
f^{\mathrm r}_{\lambda}(\phi)={\mathrm{sin}~i \over d^2} \sum_{\phi'=-\pi/2}^{\phi'=\pi/2}~\sum_{z=-h_\mathrm{r}R_\mathrm{d}}^{z=h_\mathrm{d}R_\mathrm{d}}~I^{\mathrm r}_{\lambda}(x,y)~(R_\mathrm{d}~\Delta\phi')~\mathrm{cos}~\phi'~\Delta z~O(x,y,z,\phi),
\end{equation}
where
$d$ is the distance to the system,
$i$ is the orbital inclination, and
$\phi'=\phi+\mathrm{tan}^{-1}(y/x)$
is the phase-modified disk azimuth angle.
The Cartesian coordinates of the disk rim are given by
$x=R_\mathrm{d}~\mathrm{cos}~(\phi'-\phi)$ and
$y=R_\mathrm{d}~\mathrm{sin}~(\phi'-\phi)$.

The disk surface and rim are assumed to radiate as blackbodies, with limb-darkened specific intensities given by:
\begin{equation}
I^{\mathrm s}_{\lambda}(x,y)=B_{\lambda}(T_\mathrm{d}S_\mathrm{x,y}[R_\mathrm{d}/\sqrt{x^2+y^2}]^{\alpha})~(1-\mu_{\lambda}+\mu_{\lambda} \mathrm{cos}~\gamma),
\end{equation}
\begin{equation}
I^{\mathrm r}_{\lambda}(x,y)=B_{\lambda}(T_\mathrm{d}S_\mathrm{x,y})~(1-\mu_{\lambda}+\mu_{\lambda}~\mathrm{sin}~i~\mathrm{cos}~\phi'),
\end{equation}
\noindent
where
$\mathrm{cos}~\gamma=\mathrm{cos}~\beta~\mathrm{cos}~i-\mathrm{sin}~i~\mathrm{sin}~\beta~\mathrm{cos}~\phi'$
is the cosine of the angle between the line-of-sight and the normal to the disk surface, $\beta=\mathrm{tan}^{-1}(h_\mathrm{r})$ is the half-angle of the
disk flare, and $\mu_{\lambda}$ are the linear limb-darkening coefficients.
The parameter $S_\mathrm{x,y}$ modifies the outer disk temperature in the vicinity of the intersection of the mass transfer stream and the disk periphery ($x_\mathrm{s},y_\mathrm{s}$) by the factor $\chi_\mathrm{s}$ as follows:
$$S_\mathrm{x,y} = \left\{ \begin{array}{cc}
\chi_\mathrm{s} & \mbox{ when } \hspace{4mm} \sqrt{(x-x_\mathrm{s})^2+(y-y_\mathrm{s})^2} < R_\mathrm{spot}\\
1 & \mbox{ otherwise.}
\end{array} \right.$$

The fluxes for the secondary star and the white dwarf primary are computed as follows. For the secondary star we have:
\begin{equation}
f^2_{\lambda}(\phi)={1 \over d^2} \sum B_{\lambda}(T_2) [1-\mu_{\lambda}+\mu_{\lambda} \mathrm{cos}~\xi(\phi)]~\mathrm{cos}~\xi(\phi)~\Delta A,
\end{equation}
\noindent
where $\Delta A$ is an element of area on the surface of the tidally-distorted secondary star with a normal vector inclined at an angle $\xi$ to our line of sight.
The white dwarf primary is assumed to be a spherical blackbody of temperature $T_1$ and radius $R_1$. Its integrated flux is given by:
\begin{equation}
f^1_{\lambda}(\phi)={1 \over d^2} B_{\lambda}(T_1)~(1-\mu_{\lambda}/3)~Q(R_\mathrm{in})~\pi R_1^2~O(0,0,0,\phi),
\end{equation}
where the parameter $Q$ applies a correction depending on whether the lower hemisphere of the white dwarf is shadowed by the inner edge of the disk, $R_\mathrm{in}$, as follows:
$$Q(R_\mathrm{in}) = \left\{ \begin{array}{cc}
{1+\mathrm{cos}~i \over 2}  & \mbox{ when } \hspace{4mm} R_\mathrm{in}=R_1\\
1 & \mbox{ when } \hspace{4mm} R_\mathrm{in} > R_1.
\end{array} \right.$$

Finally, the normalized light curves, independent of the distance to the system, are given by:

\begin{equation}
l_\lambda(\phi)={f_\lambda^{\mathrm s}(\phi)+f_\lambda^{\mathrm r}(\phi)+f_\lambda^{1}(\phi)+f_\lambda^{2}(\phi) \over f_\lambda^{\mathrm s}(\phi_\mathrm{n})+f_\lambda^{\mathrm r}(\phi_\mathrm{n})+f_\lambda^{1}(\phi_\mathrm{n})+f_\lambda^{2}(\phi_\mathrm{n})},
\end{equation}
where a normalization phase of $\phi_\mathrm{n}=0.2$ has been adopted to
represent the out-of-eclipse light level.

\subsection{Input Parameters}

The input parameters required by the model are summarized in Table~5.
They can be divided into two general categories:
fixed parameters and fitting parameters. The fixed parameters
are those that are either known, a priori, such as the
orbital period, or parameters whose
values can be estimated using assumptions implicit in the model.
The fitting parameters, on the other hand, cannot be
specified initially, and are varied during the fitting procedure.

Since the orbital period and eclipse width are known,
the masses and dimensions of the binary system can be computed
once the mass ratio is specified (see Paper~I).
Unfortunately, the mass ratio of V~Per has not been directly measured;
however, its value can be estimated as follows.
A lower limit on the mass ratio is established by noting that
the observed eclipse width,
$\Delta\phi=0.073$ (Wood et~al. 1992),
requires that $q~(=M_2/M_1)\grtsim0.2$ for orbital inclinations,
$i<90^{\circ}$ (Chanan et~al. 1976).
In addition, the requirement for stable mass transfer
can be used to place an upper limit on the mass ratio
given the mass of the secondary star.
The dependence of $M_2$ on $q$
is weak, and to first order the mass of the secondary star can
be expressed as a function of the orbital period alone. From
Warner (1995, eqn. 2.100) we can write $(M_2/M_{\odot})\simeq0.065P^{5/4}(hr)$.
For an orbital period of 2.57~hr, we can estimate $M_2\simeq0.2M_{\odot}$.
For secondary star masses less than $0.4~M_{\odot}$, stable mass transfer
requires $q\lessim2/3$ (e.g. Politano 1988).
Thus, the mass ratio of V~Per must lie
roughly in the range $0.2\lessim~q~\lessim0.6$.
Since for nova systems the mass of the
white dwarf is almost certainly greater than $0.5~M_{\odot}$, we can
further reduce the upper limit of plausible mass ratios to $q\simeq0.4$,
with the most likely value lying closer to $q\simeq0.2$, given that
the average white dwarf masses
in classical nova systems is believed to be of order $0.85~M_{\odot}$
(Smith \& Dhillon 1998). In order to explore the effect of the mass ratio
on our analysis, we will explore solutions for mass ratios
of both $q=0.2$ and $q=0.4$, with the former representing the most likely model.
Following the analysis described in Paper~I, the masses and dimensions of
the V~Per system have been computed for both of our representative
mass ratios, and are summarized in Table~6.

In addition to the mass ratio, the spectral type and
temperature of the secondary star
can be estimated prior to beginning the fitting procedure.
In their statistical study of the properties of cataclysmic variable
stars, Smith \& Dhillon (1998) found an empirical relation between
the orbital period and spectral type of the secondary star. For an
orbital period of 2.57~hr, the most likely spectral type is between
M4V and M5V. According to Popper (1980) a main sequence star with
a spectral type in this range is expected to have a temperature
of $\sim3300$~K. Thus, for the purposes of our study, we will adopt
$T_2=3300$~K, and not vary this parameter during the fitting procedure.

Another input parameter than can be constrained to a limited extent
is the disk rim thickness, $h_{\mathrm{r}} = h/R_{\mathrm{d}}$, where $h$ is the
height of the rim perpendicular to the plane of the disk and
$R_{\mathrm{d}}$ is the disk radius. According to the accretion disk models
of Cannizzo (2001) the pressure scale height near the
disk midplane gives
$h_{\mathrm{r}}\simeq0.01$ to a good approximation. Unfortunately,
this theoretical scale height, most often referred to by disk modelers,
is distinct from aspect ratios, $h/r\simeq0.03-0.1$, that are
measured by eclipse mapping
techniques, and relevant to our study. In particular, Cannizzo \& Wheeler
(1984) carried out vertical integrations of accretion disk structure and
showed that, even under standard $\alpha$-disk theory, the disk photosphere
can lie at $\sim~3-4$ pressure scale heights.
According to Warner (1995, eqn. 2.52), similar studies by
Meyer \& Meyer-Hofmeister (1982) and Smak (1992) show that the
rim height can be approximated by $h_{\mathrm{r}}\simeq0.038\dot M_{16}^{3/20}$.
For plausible mass accretion rates between $\sim10^{16}$~g~s$^{-1}$ and
$\sim10^{18}$~g~s$^{-1}$,
we estimate rim heights of in the range of $h_{\mathrm{r}}\sim0.04$ to
$h_{\mathrm{r}}\sim0.08$, which is consistent with our earlier estimate. 
To cover the range, we consider specific rim heights of both
$h_{\mathrm{r}}=0.04$ and $h_{\mathrm{r}}=0.08$ in our models of V~Per.
For comparisons we have also considered
a $h_{\mathrm{r}}=0.0$ ``flat disk" case for comparison with previous studies.
The values of the remaining parameters, the
``fitting parameters", which include the radial
disk temperature parameter $\alpha$,
and the temperatures of the disk rim,
bright spot, and component stars, are constrained through the model fit.

\section{Light-Curve Fitting}

To begin the fitting procedure, we select plausible ranges of values for the
five fitting parameters, indicated in Table~5 as ``variable".
Then each parameter is varied one at a time
through a suitable range while the remaining parameters
are held constant. In this way a 5-dimensional parameter space of
models is explored for plausible solutions.
The goodness of fit at each stage is determined using a standard
$\chi^2$ test.
The deviations between the model and the data are determined for
each of the four colors, and the final $\chi^2$ statistic is
determined by weighting each of the colors equally. Since we model the
light curve near eclipse only, the calculation of $\chi^2$ has been
restricted to orbital phases between $\phi=-$0.12 and $\phi=0.12$,
which corresponds roughly
to the time just prior to the
onset of disk ingress to the completion of bright-spot egress.
The range over which each parameter is varied has been chosen to be
sufficiently large so that all plausible model solutions were explored.
A thorough sampling required eight values to be considered
for each fitting parameter. Thus, a total of 32,768~($=8^5$) models
were computed for each mass ratio and rim thickness considered.
The best-fitting solutions for our grid of models are given Table~7.

In their earlier monochromatic (``white light") eclipse-mapping study of
V~Per, Wood et~al. (1992) found that the brightness temperature
gradient of the accretion disk could be made consistent with a
steady-state ($\alpha=0.75$) temperature profile only if the inner
edge of the accretion disk was of order $R_{\mathrm{in}}\sim0.2~R_{\mathrm{L1}}$,
where $R_{\mathrm{L1}}$ is the distance from the center of the white dwarf
to the inner Lagrangian point.
These authors pointed out that such a situation might arise if the
inner accretion disk was disrupted by the white dwarf's magnetic field,
as is believed to be the case in intermediate polar systems
(e.g. Warner 1995). To explore this possibility further, we have
computed an additional set of eclipse models where the inner disk radius
has been set at the value, $R_{\mathrm{in}}=0.2R_{\mathrm{L1}}$, adopted by Wood et~al. (1992).
The results of our intermediate polar models are shown in parentheses
in Table~7.

\subsection{The Model Solution Distributions}

Exploring the matrix of possible solutions reveals that many
combinations of parameters produce good fits to the data.
The goal is to determine which particular combination of parameters
best represents the true model for V~Per. The best model fits are
characterized by a reduced chi-square, $\chi_\nu^2\simeq1.5$; however,
a wide range of parameters will give comparable fits to the
light curves.
To explore the
range of plausible solutions,
frequency distributions for the top 100 best-fitting model solutions were
constructed for each of the fitting parameters.
Figures 3--5 show
the frequency distributions of the five fitting parameters,
together with the the model $B-V$ color,
for our preferred mass ratio of $q=0.20$.
Figure~3 shows the distributions for our rimless model with $h_{\mathrm{r}}=0.0$,
with Figures~4 and~5 showing the distributions for $h_{\mathrm{r}}=0.04$, and $h_{\mathrm{r}}=0.08$,
respectively.
The mean and standard deviation of each parameter distribution, along with
the mean $\chi_\nu^2$ for each model,
are summarized in Tables~8 and~9 for the standard and truncated inner disk
models, respectively.
The mean $\chi_\nu^2$ of the top 100 model solutions is significantly higher in
the rimless disk models, but still reflects acceptable fits to the data.
As long as the number of model solutions included in the distributions
is sufficiently small so that $<$$\chi_\nu^2$$>$~$\lessim1.8$
(which represents a plausible fit to the data),
the distribution means do not depend strongly
on the precise number of model solutions included in the distributions.

The frequency distributions reveal how well a given parameter
is constrained by the model fit. If a parameter is tightly constrained by
the data, its frequency distribution will be narrow and centered on its
optimum value.
Thus, the mean and standard deviation of each distribution not only provide
an estimate of the optimum value for a specific parameter, but also
a quantitative assessment of its uncertainty.
The most tightly constrained parameter is clearly the
disk radius, $R_{\mathrm{d}}$,
where a radius of $\sim70-75$\% of the distance to the inner Lagrangian
point is required by virtually all the models.
The value of this parameter is primarily
determined by the phase width where the disk eclipse begins and ends,
and is relatively insensitive to other system parameters
(e.g. Sulkanen et~al. 1981). The light curves do not exhibit any significant
asymmetry of the eclipse profile, and there is no prominent``hump" prior to
eclipse, so it is not surprising that
none of the models considered required significant radiation from the
bright spot in the system. Furthermore, the eclipse profile,
which lacks the sharp transitions on ingress and egress
characteristic of the white dwarf eclipse
suggest that the white dwarf star does not contribute significantly
to the system light. Thus, as expected,
none of the models we considered required a
luminous (hot) white dwarf. As an example,
Figure~6 shows the best-fitting $q=0.20$ and $h_{\mathrm{r}}=0.08$ model
plotted together with the phased eclipse data. The best fitting models
for $q=0.4$, and for other rim thicknesses fit the data almost as well,
and cannot be ruled out based on the fit alone.

\section{Discussion}

\subsection{Disk Structure}

Monochromatic eclipses of V~Per have been studied by Wood et~al. (1992)
using a maximum entropy eclipse-mapping technique.
A principal conclusion of their study was that the brightness
temperature distribution of the accretion
disk was flatter than that expected for a steady-state disk.
These authors showed, however, that 
if the inner disk in V~Per were disrupted out to
$R_{\mathrm{in}}\sim 0.15-0.25R_{\mathrm{L1}}$,
as might be expected if V~Per harbors
a magnetic white dwarf, then a brightness temperature gradient
very close to that expected for a steady-state disk
could be achieved.

Our multi-color eclipse simulations are qualitatively in agreement
with the results of Wood et~al. (1992).
Referring to Table~7, we see that for a given mass ratio and rim thickness,
the intermediate polar models are generally characterized by slightly
larger values
of the disk temperature parameter, $\alpha$. However, in our case,
{\it only models with
a disk rim produced a best-fitting value of $\alpha$ close to the
steady-state value.} Only two cases -- 
the $q=0.2$, $h_{\mathrm{r}}=0.08$ standard model and the $q=0.4$,
$h_{\mathrm{r}}=0.08$ intermediate polar model -- resulted in a
value of $\alpha=0.75$. It is noteworthy that for a given rim height
the value of $\alpha$ is systematically higher in all of the intermediate
polar models compared with the standard models,
underscoring the synergistic effect between the rim and
the truncated inner disk in achieving a steady-state disk solution.
For the orbital inclination of 85.4$^{\circ}$ required by our $q=0.2$ model,
the central region of the accretion disk, including the white dwarf, is 
just barely visible with a rim thickness of $h_{\mathrm{r}}=0.08$.
Any ``thick disk" models characterized by
rim height $h_{\mathrm{r}}>0.08$ would result in complete shadowing
of the white dwarf and part of the disk at all orbital phases,
and thus would be insensitive to the white dwarf's luminosity.

Wood et~al. also considered an alternative model with a disk rim,
and found that their rim model did not result in a significantly
steeper brightness temperature gradient. However,
the rim model they considered was considerably
less sophisticated than ours. In particular, their
models employed a ``dark" disk rim solely to study the
effect produced by shadowing of the disk by the rim.
The effect of the rim's contribution to the overall disk light
was not included in their models.
Furthermore, rim heights of only 1--2\% of the
disk radius were considered, which required inclination angles,
$i>89^{\circ}$, to achieve significant rim shadowing.

As a test of the importance of a bright rim versus a dark rim like that
considered by Wood et al., we ran a series of models with our standard flared disk,
but with the rim brightness switched off. These ``dark rim" models generally
produced the same best-fit solutions as our standard models, but with
one difference:
a significantly lower value of the disk radial temperature parameter, $\alpha$.
Specifically, for our four standard models with a rim
($q=0.2, h_\mathrm{r}=0.04; q=0.2, h_\mathrm{r}=0.08; q=0.4, h_\mathrm{r}=0.04; q=0.4, h_\mathrm{r}=0.08$),
we found values of $\alpha=0.15, 0.45, 0.25, \mathrm{and}~0.35$,
respectively, compared with
the values of $\alpha=0.35, 0.75, 0.25, \mathrm{and}~0.45$ from our bright rim
models (see Table~7). The only case where the bright rim apparently
made no difference was in the model characterized by $q=0.4, h_\mathrm{r}=0.04$.
However, in this case, the lower inclination and modest rim height make the
contribution of the rim to the total disk light
the smallest of all models considered.

The importance of a bright disk rim can be best visualized by referring
to Figures~7 and~8, which show the system geometry at mid-eclipse.
At high inclinations the disk rim makes a significant contribution to
the total disk light.
Since the disk rim (excluding the bright spot) has a nearly constant
surface brightness, the eclipse of this rimmed disk will be more
{\bf V}-shaped than the eclipse of a flat disk,
even in the case where the disk has a steady-state ($\alpha=0.75$) 
radial temperature profile (e.g., see the eclipse simulations
and discussion of DW~UMa by Knigge et al. 2000, 2004).
As mentioned earlier, Smak (1994) has shown that modeling such eclipses
under the assumption that the disk is flat (as in eclipse mapping)
leads to the spurious conclusion that the disk temperature profile is shallower
that $\alpha=0.75$.
In the case of V~Per, an additional effect is that
for both mass ratios considered,
the accretion disk is never completely eclipsed. In models with
a disk rim, the contribution of the rim
to the total system light at mid-eclipse can be significant. In order to
produce an eclipse with the observed depth (in the {\it normalized}
light curves), the inclusion of a bright outer disk rim
requires that the eclipsed inner disk be sufficiently bright to compensate,
i.e. to keep the
ratio of the mid-eclipse to out-of-eclipse system brightness unchanged.
For a given outer disk temperature, this can only be
accomplished by having a luminous white dwarf, or by having
a steep disk temperature profile, or both. These two possibilities produce
very different ingress and egress eclipse profiles, and can therefore
be distinguished by the model fit. In the case of V~Per, the steep ingress
and egress characteristic of a significant white dwarf contribution to the
overall luminosity is not observed, leaving a steep disk temperature gradient
as the only viable alternative.

\subsection{Correlations Between Model Parameters}

Further insights into the the effect that the various input parameters
have on acceptable model solutions can be found by exploring correlations
between the parameters. We begin by considering an expanded set of model
solutions defined by $\chi_\nu^2<2.0$, which allows potential correlations
to be studied over an expanded range in parameter space. Correlations between
pairs of parameters are explored by allowing these parameters to vary while
holding the remaining parameters fixed at their optimum values. For a total
of five fitting parameters, there are a total of 10 such pairings. As
a representative example,
the correlations for the best-fitting $q=0.2$, $h_{\mathrm{r}}=0.04$
model are shown in Figure~9.

As in previous studies of GY~Cnc and EX~Dra (Papers~I and~II),
the most highly correlated
pairs involve $R_{\mathrm{d}}$, $T_{\mathrm{d}}$, and $\alpha$, which together determine the
disk luminosity. There is a major difference in the correlations found
in the present study, however. In the cases of GY~Cnc and EX~Dra, which
are longer orbital period systems with larger mass ratios, the accretion
disks were entirely occulted at mid-eclipse. Thus, the depth of eclipse
fixes the disk luminosity. For a constant disk luminosity, the radial
temperature parameter $\alpha$ should be negatively correlated with the
temperature at the disk periphery, $T_{\mathrm{d}}$, and the disk radius, $R_{\mathrm{d}}$,
as was indeed observed for these two systems. In the case of V~Per,
however, the situation is more complicated as described in the previous
section. Because the disk is never completely obscured,
the eclipse depth no longer uniquely defines the disk luminosity.
The normalized eclipse depth can be achieved with a relatively
low-luminosity
disk having a relatively shallow radial temperature gradient, or with
a disk having a bright rim (relatively high temperature and/or vertical extent)
and steep temperature gradient (large $\alpha$).
As the contribution of the disk rim is increased, the inner disk
luminosity must increase to compensate, requiring a steeper temperature
gradient and a higher value of $\alpha$.
Thus, for a given disk radius and rim thickness,
$\alpha$ should be {\it positively} correlated with the outer
disk (rim) temperature, $T_{\mathrm{d}}$, as observed in the case of V~Per, and
shown in Figure~9.
In view of these considerations, it
is not surprising that the $q=0.40$, $h_{\mathrm{r}}=0.00$ and
$q=0.20$, $h_{\mathrm{r}}=0.08$ models
yield the least luminous and most luminous disks, respectively.

In principle, a smaller disk with the appropriate luminosity
that is completely eclipsed could produce the observed eclipse depth,
however, the shape of the eclipse profile, not just its depth is
important for a good model fit. In the case of V~Per, the maximum width
of the eclipse profile (as defined by the onset of ingress and the termination
of egress) requires a relatively large
disk radius that approaches the tidal limit
(see Figure 6). 

\subsection{Model Constraints from the Disk Luminosity}

Although a range of models can produce acceptable fits to the data,
as noted above, these models differ in one important respect:
they represent significantly different disk luminosities.
Referring to our best-fitting model solutions (Table~7), we are reminded
that rimless models are characterized by flat temperature profiles
(small $\alpha$).
For a given outer disk radius and temperature, disks with relatively
flat temperature gradients will be considerably
less luminous than steady-state (rimmed) disks characterized by $\alpha=0.75$.
Thus, a potential way to constrain the models further is to favor
those models that produce the expected disk luminosity, and absolute
magnitude for V~Per.
The disk luminosities can be explored through two independent approaches:
directly from our model fluxes, and from an estimate of the mass
accretion rate inferred from our models.

In the first approach, we can
compare the ratio of our model $V$ fluxes for the disk and secondary star
as viewed from an inclination angle of $i=0$ (i.e. ``face-on"). This
quantity, $\eta_\mathrm{d,s}=F_\mathrm{V}^\mathrm{d}(i=0)/F_\mathrm{V}^\mathrm{s}(i=0)$,
is given in Table~10 for our best-fitting models. If we ignore limb-darkening
and assume a
spherical secondary star and a flat circular disk, then the luminosity
ratio is approximately equal to
half the model flux ratio (a flat disk has half the surface area
of a sphere). Thus,
$L_\mathrm{V}^\mathrm{d}/L_\mathrm{V}^\mathrm{s}\simeq\eta_\mathrm{d,s}/2$, and
an estimate of the secondary
star's luminosity will yield a crude estimate for the disk luminosity.
In reality, this estimate will be a slight underestimate of the disk luminosity
because we have neglected the relatively small contribution of the rim
to the overall disk luminosity.
As an example, the flux ratio of our best-fitting $q=0.2, h_\mathrm{r}=0.08$ model
is $\eta_\mathrm{d,s}=3.18\times10^3$,
which represents a luminosity ratio of $\sim1.59\times10^3$. Thus,
we estimate the
disk in this model to be $\sim8.0$ mag brighter than the secondary star.
An estimate of $M_\mathrm{V}(2)$
can be found given the secondary star's spectral type,
which we estimated earlier to be $\sim$M4V--M5V.
For a main-sequence secondary star of radius
$R_2=0.24$~R$_{\odot}$ (Table~6) and spectral type M4V--M5V,
we find $12.4<M_{\mathrm{V}}(2)<12.8$ (Popper 1980). Thus, for this model,
we estimate an absolute visual magnitude for the accretion disk, and in
effect for the V~Per system (the white dwarf and secondary star
contribute little $V$ light compared to the disk),
to be $M_\mathrm{V}\simeq4.6$.
Although somewhat bright for a $P=2.6$~hr system, this
value is certainly plausible for V~Per, and is
typical for the absolute visual magnitude of nova remnants
(e.g. Warner 1995).

As a check on this result, we can also make a rough
estimate the disk luminosity
from an estimate of the mass transfer rate coupled with the $\dot M$ vs.
$M_\mathrm{V}$ relation from Tylenda (1981). We begin
with a well-known expression for the effective temperature of the
accretion disk (e.g. Pringle 1981):
\begin{equation}
\sigma T^4(r) = {3 G M_1\dot M \over 8\pi r^3}~\left[1 - \sqrt{R_\mathrm{in} \over r}\right],
\end{equation}
where $R_\mathrm{in}$ is the inner radius of the disk.
By adopting the model radius and temperature of the outer disk (rim),
we can compute the mass accretion rate for a given model. Specifically,
\begin{equation}
\dot M = {8 \pi R_\mathrm{d}^3 \sigma T_\mathrm{d}^4 \over 3 G M_1}\left[1 - \sqrt{R_\mathrm{in} \over r}\right]^{-1} \mathrm{gm~s}^{-1}.
\end{equation}
Mass transfer rates ($\dot M_{17} = \dot M/10^{17}~\mathrm{g~s}^{-1}$)
for our best-fitting models are given in Table~10.
For the specific case of the $q=0.2, h_\mathrm{r}=0.08$ model with $\alpha=0.75$,
the mass-transfer rate can be converted to an equivalent $M_\mathrm{V}$ through an
extrapolation of the relation given in Table~1 of Tylenda (1981).\footnote{
It has
been necessary to apply a small correction ($\sim0.8$~mag) to the resulting
absolute magnitude to account for the difference in $M_1$ and $R_\mathrm{d}$
between Tylenda's model ($M_1=1M_{\odot},
R_\mathrm{d}=5\times10^{10}~\mathrm{cm}$) and our model.
Following Warner (1995),
the correction for $R_\mathrm{d}$ has been estimated from the models of
Wade (1984).} We find a value of $M_\mathrm{V}\simeq4.3$, which is
consistent with the value of $M_\mathrm{V}\simeq4.6$
estimated earlier from the disk-to-secondary star luminosity ratio.

The absolute magnitudes
and accretion rates
of systems with flat accretion disks
($M_\mathrm{V}\simeq7-8$ and
$\dot M_{17}\simeq1.5-2.5$)
are more typical of dwarf nova disks at minimum light.
Such feeble disks with
their shallow temperature gradients ($\alpha=0.15$) are unlikely
to be seen in nova and nova-like systems with stable accretion
unless a significant amount of mass and accretion
energy is being lost from the inner disk, for example, by driving
a wind. In such a case, the
mass accretion rate in the outer disk
could possibly be increased without necessarily
producing a luminous disk with a steep temperature profile.

Since V~Per does not exhibit dwarf nova eruptions,
a further constrain on possible models for V~Per is provided
by the requirement that the mass transfer rate exceed the citical
value for stable accretion, $\dot M_\mathrm{crit}$.
A convenient expression for the critical mass transfer rate
is given by
$\dot M_\mathrm{crit}\simeq10^{16} R_\mathrm{10}^{21/8} M_1^{-7/8}$~gm~s$^{-1}$, where $R_\mathrm{10}$ is the radius of the disk in units of
$10^{10}$~cm (e.g. see Shafter et al. 1986, and references therein).
Values of $\dot M_\mathrm{crit}$ for our best-fitting models
are given in Table~10. The flat disk models
have mass accretion rates that are significantly below the critical value,
while only the thick disk models have rates clearly in excess of the
critical value.

To minimize the number of fitting parameters in our models, we
have set the disk rim temperature equal to the minimum
temperature in the outer disk, $T_\mathrm{d}$.
It seems possible, however, that this assumption
overestimates the value of $T_\mathrm{d}$ used to compute the disk
temperature profile since energy input from the mass-transfer stream
is likely to raise the temperature of the disk rim over that appropriate
for the outer disk region.
If so, then our estimates of mass transfer rates
and disk luminosities are best considered upper limits.
In this case, the flat disk models with shallow temperature gradients
become even more problematic, and
our conclusion favoring the high-mass-accretion-rate,
high-luminosity, thick-disk model for the V~Per system is strengthened.

\subsection{The Distance to V~Per}

The distance to V~Per can be estimated
by comparing the observed brightness at mid-eclipse with
estimates of the absolute magnitude of the secondary star,
and the fraction of light it contributes at mid-eclipse.
The fraction of light contributed by the secondary star
at mid-eclipse,
$f_2=f^\mathrm{s}_\mathrm{V}(\phi=0)/f^\mathrm{tot}_\mathrm{V}(\phi=0)$,
depends on the model considered, and in particular on
the properties of the disk rim. Values of $f_2$ for each model
are given in Table~10.
In the rimless models, the disk
contributes relatively little to the light at mid-eclipse compared
with the models with rims.
According to our best-fitting solutions for models with
a thick rim, the secondary
star contributes $\sim1.0$\% of the light at mid-eclipse,
while in the rimless models, where the disk luminosity is relatively
low, the secondary star contributes a substantial
$\sim20$\% of the mid-eclipse light.
Given that $m_\mathrm{V}=19.34$ (see Table~3) at mid-eclipse,
we estimate the apparent distance moduli $(m-M)_\mathrm{V}$ given in
Table~10.

The interstellar extinction along the line of sight can be estimated
from the color excess obtained from
a comparison of the observed and model $B-V$ colors.
V~Per has an observed color of $B-V\simeq0.6\pm0.2$ outside of eclipse,
where much of the uncertainty is due to the limited phase coverage of our
outside eclipse $B$-band data.
Assuming a ratio of total-to-selective extinction of $R_\mathrm{V}=3$ yields
the visual extinctions, $A_\mathrm{V}$, along the line of sight to V~per
given in Table~10.
Our models with a rim have
colors outside of eclipse that are significantly bluer than the
colors of the rimless models, leading to larger values of $A_\mathrm{V}$.
Correcting the apparent distance moduli yields
the model-dependent distance estimates given in Table~10.

As a consistency check,
the distance to V~Per can also be determined from a comparison
of the out-of-eclipse magnitude with an estimate of the system's
absolute magnitude corrected for inclination. Warner (1987, 1995) gives
a correction factor for the conversion of absolute magnitude to
{\it apparent\/} absolute magnitude of
$\Delta M_\mathrm{V}(i)=-2.5~\mathrm{log} \left[(1+1.5~\mathrm{cos}~i)~\mathrm{cos}~i \right]$;
however, this correction does not take into account any contribution
from the disk rim.\footnote{The expression given in Warner (1995) corrects
an error in the Warner (1987) formula, however the sign is
incorrect for his stated conversion of apparent absolute magnitude to
absolute magnitude.}
Fortunately, the ratio of rim-to-disk light
($\eta_\mathrm{r,d}$) is given by our models (see Table~10).
In the case of our favored $q=0.2, h_\mathrm{r}=0.08$ model,
the rim contributes $\sim50$\% of the disk light. Applying these corrections for
inclination, taking
$M_\mathrm{V}=4.6$ for the $q=0.2, h_\mathrm{r}=0.08$ model,
and adopting $m_\mathrm{V}=18.1$ for V~Per out of eclipse (see Table~3),
yields a reddening-corrected distance modulus of $(m-M)_\mathrm{V}=9.6$,
and a corresponding distance of $\sim820$~pc.
Given the uncertainties,
this value is consistent with our estimate of 1~kpc
based on the properties of the secondary star.

In summary,
we estimate that V~Per lies
at a distance of $\sim1.0$~kpc,
in our favored scenario where the disk
has a significant rim that contributes to the light at mid-eclipse.
In contrast,
our rimless models, which predict that a significantly larger fraction
of the light at mid-eclipse arises from the secondary star, suggest
that V~Per lies at a distance of only $\sim0.5$~kpc. 
In both cases, a significant component of the uncertainty
in our distance estimates
arises from uncertainties in the visual extinction.
Overall, our favored
distance estimate is consistent with the value of $\sim1$~kpc
derived by Shafter (1997) from a comparison of the apparent magnitude
at maximum light ($m_{\mathrm{max}}\simeq4$), an assumed absolute magnitude
at maximum light, $M_\mathrm{V}=-7.5$, and a visual extinction of $A_\mathrm{V}=1.4$
estimated from the reddening maps of Neckel et~al. (1980).

\section{Conclusions}

We have performed the first multicolor eclipse study of the classical
nova V~Per using a parameter-fitting eclipse code to model the
light curve. Our study has explored
a variety of assumptions regarding the structure of the accretion disk,
including models with a large inner disk radius, as might be expected
if V~Per has a strongly magnetic white dwarf, and models that include
a disk rim. In agreement with the earlier monochromatic maximum
entropy eclipse mapping study of V~Per by
Wood et~al. (1992), we find that a flat accretion disk that extends
down to the surface of the white dwarf must have a brightness temperature
profile that is significantly flatter than that expected for a
steady-state disk.
In an attempt to resolve
the discrepancy, Wood et~al. explored models where the inner disk
was disrupted, as might be expected if V~Per harbors a magnetic white dwarf.
They found that temperature profiles consistent with the steady-state case
could be produced in
models with an increased inner disk radius, $R_{\mathrm{in}}\simeq0.2R_{\mathrm{L1}}$.

Although the brightness temperature profiles of our best-fitting
models become steeper
when the inner disk is truncated out to $R_{\mathrm{in}}=0.2R_{\mathrm{L1}}$,
unlike Wood et~al., we were unable to reproduce a flat-disk
model with $\alpha\simeq0.75$
with a truncated inner disk alone. However, our models that included a
disk rim were characterized by considerably steeper brightness
temperature profiles, particularly in our ``thick disk"
($h_{\mathrm{r}}=0.08$) case where we found $\alpha=0.75$
gives the best fit to the data.
Although acceptable fits for models with rims can be found with smaller
values of $\alpha$ and lower outer disk temperatures (faint rims),
the {\it only\/} way to achieve $\alpha=0.75$ is with a model having a
bright rim. Furthermore, only the bright rim models give accretion rates
and absolute magnitudes that are typical of nova systems with
stable accretion. 
By extrapolating our result for V~Per,
we confirm the conclusion reached by Smak (1994)
from his generic eclipse simulations,
that prior attempts to analyze cataclysmic
variable eclipses with flat-accretion-disk models have significantly
underestimated the brightness temperature gradients of the accretion disks.

In summary, regardless of the adopted mass ratio or rim thickness,
we find that V~Per is most likely characterized
by a large accretion disk (extending to near its tidal radius), characterized
by $R_{\mathrm{d}}\simeq0.75R_{\mathrm{L1}}$, with a temperature $T_{\mathrm{d}}\simeq6000-10000$K
at its outer edge. No evidence was found for an unusually
luminous (hot) white dwarf
or bright spot in the system.
However, if the disk is in a steady-state, optically-thick
configuration, we have shown that the disk must possess a significant rim.
As in V~Per, it is also likely that
other short-period cataclysmic variable stars with mass ratios, $q\lessim0.5$,
will also potentially harbor accretion disks that are not fully occulted
at mid-eclipse. As shown in the present study, it is imperative that any
attempts to model the accretion disks in these systems should take into account
the effect of a disk rim.


\acknowledgments
We thank the staff of the Steward Observatory for granting time on the
Bok 2.3~m reflector and Karl Gordon for assistance with the observations.
We also thank John Cannizzo and Ted Daub for reading an earlier version of
the manuscript and for subsequent discussion,
Jerry Orosz for computing pictorial representations of the
system geometry near mid-eclipse, and the referee, I. B. B\'{\i}r\'o, for
a thorough report that improved our presentation. We are also grateful to
Christian Knigge and Jean-Pierre Lasota
for bringing an important reference
omitted in the original manuscript to our attention.
Our eclipse model is based in part on
occultation kernel subroutines kindly provided by K. Horne.

\clearpage

\begin{deluxetable}{lcc}
\tablenum{1}
\tablecaption{CVs in the Period Gap}
\tablewidth{0pt}
\tablehead{
 \colhead{Name}  & \colhead{Type} & \colhead{Orbital Period (hr)}
}
\startdata
V516 Pup            & AM     & 2.29 \\
SDSS J205017-053627 & NL     & 2.30 \\
QS Tel              & AM     & 2.33 \\
DD Cir              & N, DQ  & 2.34 \\
NY Ser              & SU     & 2.35 \\
SDSS J080908+381406 & NL     & 2.38 \\
V348 Pup            & DQ     & 2.44 \\
MN Dra              & SU     & 2.50 \\
RX J1554.2+2721     & AM     & 2.53 \\
V Per               & N      & 2.57 \\
V795 Her            & DQ     & 2.60 \\
1RXS J052430+424449 & AM, DQ & 2.62 \\
V349 Pav            & AM     & 2.66 \\
QU Vul              & N      & 2.68 \\
SDSS J075240+362823 & AM?    & 2.70 \\
\enddata
\end{deluxetable}

\clearpage

\begin{deluxetable}{lcccc}
\tablenum{2}
\tablecaption{Summary of Observations}
\tablewidth{0pt}
\tablehead{
 & \colhead{UT Time}  & \colhead{Time Resolution\tablenotemark{a}} & \colhead{Number of} & \\
\colhead{UT Date} & \colhead{(start of observations)} & \colhead{(sec)} & \colhead{Exposures} & \colhead{Filter}
} 
\startdata
2002 Oct 10 & 4:57:00 & 41.5 & 371 & V \\
2002 Oct 10 &10:34:00 & 41.5 & 151 & V \\
2002 Oct 11 & 7:05:00 & 57.0 & 248 & B \\
2002 Oct 29 & 4:33:00 & 56.1 &  74 & R \\
2002 Oct 29 & 7:00:00 & 56.1 &  74 & R \\
2002 Oct 29 & 9:11:38 & 56.2 & 121 & I \\
2002 Dec 09 & 2:20:35 & 56.4 & 111 & I \\
2002 Dec 09 & 5:22:00 & 56.4 &  77 & B \\
2002 Dec 09 & 7:32:28 & 56.4 &  96 & I \\
2002 Dec 11 & 5:17:00 & 56.4 & 132 & R \\
\enddata
\tablenotetext{a}{Mean time interval between exposures (Integration time plus readout time)}
\end{deluxetable}

\clearpage

\begin{deluxetable}{lccc}
\tablenum{3}
\tablecaption{Mean Magnitudes and Colors}
\tablewidth{0pt}
\tablehead{
& \colhead{Effective} & &  \\
\colhead{Photometric} & \colhead{Wavelength} & & \\
\colhead{Parameter} & \colhead{(\AA)} & \colhead{Out of Eclipse\tablenotemark{a}} & \colhead{Primary Minimum}
} 
\startdata
$B$\dots\dots\dots & 4375 & $18.7\pm0.1$ & $20.05\pm0.09$ \\
$V$\dots\dots\dots & 5460 & $18.1\pm0.1$ & $19.34\pm0.08$ \\
$R$\dots\dots\dots & 6543 & $18.1\pm0.2$ & $19.13\pm0.31$ \\
$I$\dots\dots\dots & 7992 & $17.5\pm0.1$ & $18.61\pm0.17$ \\
\\
$B-V$\dots & \dots & $0.6\pm0.2$ & $0.71\pm0.12$ \\
$V-R$\dots & \dots & $0.0\pm0.3$ & $0.21\pm0.32$ \\
$V-I$\dots & \dots & $0.6\pm0.2$ & $0.73\pm0.19$ \\

\enddata
\tablenotetext{a}{The mean magnitude of the light curve excluding
orbital phases in the range: $-0.15<\phi<0.15$.}
\end{deluxetable}

\clearpage

\begin{deluxetable}{lccrc}
\tablenum{4}
\tablecaption{Eclipse Timings}
\tablewidth{0pt}
\tablehead{
\colhead{HJD (mid-eclipse)} & Cycle Number & & \colhead{$O-C$} \\
\colhead{(2,400,000+)} & \colhead{$(E)$} & \colhead{Filter} & \colhead{($\times10^3$~day)} & \colhead{reference}\tablenotemark{a}
} 
\startdata
47442.8264\dots &      0. &BVR&  $ 0.438541$ & 1 \\
47442.9337\dots &      1. &BVR&  $ 0.615067$ & 1 \\
47443.7896\dots &      9. &BVR&  $-0.472726$ & 1 \\
47443.8965\dots &     10. &BVR&  $-0.696200$ & 1 \\
47444.0040\dots &     11. &BVR&  $-0.319674$ & 1 \\
47446.8969\dots &     38. &BVR&  $ 0.246525$ & 1 \\
47447.0038\dots &     39. &BVR&  $ 0.023051$ & 1 \\
47448.9321\dots &     57. &BVR&  $ 0.100518$ & 1 \\
50430.7140\dots &  27892. & V &  $ 0.099267$ & 2 \\
50430.8207\dots &  27893. & V &  $-0.324207$ & 2 \\
50481.5978\dots &  28367. & R &  $ 0.249075$ & 2 \\
50481.7052\dots &  28368. & B &  $ 0.525601$ & 2 \\
50509.6639\dots &  28629. & V &  $-0.001136$ & 2 \\
50692.8455\dots &  30339. & B &  $ 0.458173$ & 2 \\
50695.8445\dots &  30367. & R &  $ 0.000898$ & 2 \\
50696.9157\dots &  30377. & B &  $-0.033843$ & 2 \\
50696.8089\dots &  30376. & R &  $ 0.289631$ & 2 \\
50730.7668\dots &  30693. & I &  $ 0.048345$ & 2 \\
50730.8740\dots &  30694. & I &  $ 0.124871$ & 2 \\
50730.9807\dots &  30695. & I &  $-0.298603$ & 2 \\
50731.9451\dots &  30704. & V &  $-0.009870$ & 2 \\
\enddata
\end{deluxetable}
 
\clearpage

\begin{deluxetable}{lccrc}
\tablenum{4 (cont.)}
\tablecaption{Eclipse Timings}
\tablewidth{0pt}
\tablehead{
\colhead{HJD (mid-eclipse)} & Cycle Number & & \colhead{$O-C$} \\
\colhead{(2,400,000+)} & \colhead{$(E)$} & \colhead{Filter} & \colhead{($\times10^3$~day)} & \colhead{ref}\tablenotemark{a}
} 
\startdata
50757.8691\dots &  30946. & R &  $ 0.109401$ & 2 \\
50758.7253\dots &  30954. & I &  $-0.678392$ & 2 \\
50758.8331\dots &  30955. & B &  $-0.001866$ & 2 \\
50759.7971\dots &  30964. & B &  $-0.113133$ & 2 \\
50759.9041\dots &  30965. & V &  $-0.236607$ & 2 \\
52172.541\dots  &  44152. & R &  $-0.589411$ & 3 \\
52557.7580\dots &  47748. & V &  $ 0.397768$ & 4 \\
52557.8642\dots &  47749. & V &  $-0.525707$ & 4 \\
52557.9720\dots &  47750. & V &  $ 0.150819$ & 4 \\
52558.8293\dots &  47758. & B &  $ 0.463027$ & 4 \\
52558.9362\dots &  47759. & B &  $ 0.239553$ & 4 \\
52576.7186\dots &  47925. & R &  $ 0.142854$ & 4 \\
52576.8255\dots &  47926. & R &  $-0.080620$ & 4 \\
52576.9323\dots &  47927. & I &  $-0.404094$ & 4 \\
52617.6398\dots &  48307. & I &  $ 0.175752$ & 4 \\
52617.7469\dots &  48308. & B &  $ 0.152278$ & 4 \\
52617.8537\dots &  48309. & I &  $-0.171196$ & 4 \\
52619.7820\dots &  48327. & R &  $-0.093730$ & 4 \\
\enddata
\tablenotetext{a}{1.-- Shafter \& Abbott 1989; 2.-- Dahm (1997); 3.-- Katysheva et~al. 2002; 4.-- This work.}
\end{deluxetable}

\clearpage

\begin{deluxetable}{lcc}
\tablenum{5}
\tablecaption{Model Input Parameters}
\tablewidth{0pt}
\tablehead{
\colhead{Parameter} & \colhead{Definition} & \colhead{Value\tablenotemark{a}}
} 
\startdata
$P$\dots\dots & Orbital Period, $P$ & 2.57~hr \\
$q$\dots\dots & Mass ratio, $M_2/M_1$ & 0.20,0.40 \\
$\Delta\phi$\dots\dots & Eclipse phase width & 0.073\\
$i$\dots\dots & Orbital inclination & computed \\
$\alpha$\dots\dots & Disk temperature parameter & variable\\
$R_{\mathrm{d}}$\dots\dots & Disk radius & variable\\
$R_{\mathrm{in}}$\dots\dots & Inner disk radius & $R_1$, 0.2$R_{\mathrm{L1}}$\\
$h_{\mathrm{r}}$\dots\dots & Disk rim parameter& 0.00,0.04,0.08 \\
$T_{\mathrm{d}}$\dots\dots & Temperature of disk perimeter & variable\\
$R_1$\dots\dots & Radius of white dwarf & computed\\
$R_2$\dots\dots & Radius of secondary star & computed\\
$T_1$\dots\dots & Temperature of white dwarf & variable\\
$T_2$\dots\dots & Temperature of secondary star & 3300~K\\
$\chi_\mathrm{s}$\dots\dots & Bright spot temperature factor & variable \\
$R_s$\dots\dots & Bright spot radius & 0.2$R_{\mathrm{d}}$ \\
\enddata
\tablenotetext{a}{Where given explicitly,
the values are fixed and not varied during the
fitting procedure. The values of ``variable" parameters are determined
by the model during the fitting procedure, while the
``computed" values are calculated by the model for a given mass ratio.}
\end{deluxetable}

\clearpage

\begin{deluxetable}{lcc}
\tablenum{6}
\tablecaption{Binary Parameters}
\tablewidth{0pt}
\tablehead{
\colhead{Parameter} & \colhead{$q=0.20$} & \colhead{$q=0.40$}}
\startdata
$i(^\circ)$\dots\dots\dots    &   85.4 &      79.4  \\
$a(R_{\odot})$\dots\dots\dots &   0.95 &      0.80  \\
$R_{\mathrm{L1}}(a)$\dots\dots\dots    &   0.66 &      0.59  \\
$R_1(10^{-2}~R_{\odot})$\dots &   0.97 &      1.52  \\
$M_1(M_{\odot}$)\dots\dots    &   0.85 &      0.43  \\
$R_2(R_{\odot})$\dots\dots .  &   0.24 &      0.24  \\
$M_2(M_{\odot}$)\dots\dots    &   0.17 &      0.17  \\
\enddata
\end{deluxetable}

\clearpage

\begin{deluxetable}{lccccccc}
\tablenum{7}
\rotate
\tablecaption{Model Grid Solutions\tablenotemark{a}}
\tablewidth{0pt}
\tablehead{
& & \colhead{$q=0.20$} & & \colhead{} & & \colhead{$q=0.40$} & \\ 
\colhead{Parameter} & \colhead{$h_{\mathrm{r}}=0.00$} & \colhead{$h_{\mathrm{r}}=0.04$} & \colhead{$h_{\mathrm{r}}=0.08$} & \colhead{} & \colhead{$h_{\mathrm{r}}=0.00$} & \colhead{$h_{\mathrm{r}}=0.04$} & \colhead{$h_{\mathrm{r}}=0.08$}} 
\startdata
$R_{\mathrm{d}}(a)$           & 0.49 (0.40) & 0.49 (0.40) & 0.46 (0.40) & &  0.41 (0.41) & 0.44 (0.44) & 0.41 (0.38) \\
$R_{\mathrm{d}}(a)_{\mathrm{lim}}$\tablenotemark{b} & 0.50 (0.50) & 0.50 (0.50) & 0.50 (0.50) & & 0.43 (0.43) & 0.43 (0.43) & 0.43 (0.43) \\
$R_{\mathrm{d}}(R_{\mathrm{L1}})$      & 0.65 (0.60) &  0.75 (0.65) & 0.75 (0.70) & & 0.70 (0.65) & 0.75 (0.70) & 0.75 (0.75) \\
$\alpha$           & 0.15 (0.35)&  0.35 (0.55) & 0.75 (0.75) & & 0.15 (0.25) & 0.25 (0.45) & 0.45 (0.65) \\
$T_1$($10^3$~K)    & 10.0 (10.0 )&  10.0 (15.0) &  10.0 (25.0) & & 10.0 (10.0) & 10.0 (10.0) & 10.0 (15.0) \\
$T_{\mathrm{d}}$($10^3$~K)    & 6.0 (6.0) &  6.0 (6.0) &  10.0 (6.0) & & 6.0 (6.0) & 6.0 (6.0) & 8.0 (6.0)\\
$\chi_\mathrm{s}$              & 1.2 (1.2) & 1.2 (1.1) & 1.3 (1.1) & & 1.1 (1.1) & 1.3 (1.2) & 1.3 (1.2) \\
$(B-V)_{\circ}$    & 0.63 (0.56) &  0.45 (0.44) & 0.03 (0.34) & & 0.59 (0.56) & 0.49 (0.45) & 0.19 (0.33) \\
$\chi_{\nu,{\mathrm{min}}}^2$     & 1.7 (1.7) &  1.6 (1.6) &  1.5 (1.6) & & 1.6 (1.6) & 1.5 (1.6) & 1.5 (1.5) \\
\enddata
\tablenotetext{a}{Values in parenthesis represent intermediate polar models
where the inner accretion disk has been truncated at $R_{\mathrm{d}}(R_{\mathrm{L1}})=0.2$.}
\tablenotetext{b}{The limiting tidal radius of the disk computed using
eqn~(2.61) of Warner (1995).}
\end{deluxetable}

\clearpage

\begin{deluxetable}{lccccccc}
\tablenum{8}
\rotate
\tablecaption{Standard Model Distribution Mean Solutions}
\tablewidth{0pt}
\tablehead{
& & \colhead{$q=0.20$} & & \colhead{} & & \colhead{$q=0.40$} & \\ 
\colhead{Parameter} & \colhead{$h_{\mathrm{r}}=0.00$} & \colhead{$h_{\mathrm{r}}=0.04$} & \colhead{$h_{\mathrm{r}}=0.08$} & \colhead{} & \colhead{$h_{\mathrm{r}}=0.00$} & \colhead{$h_{\mathrm{r}}=0.04$} & \colhead{$h_{\mathrm{r}}=0.08$}} 
\startdata
$<\chi_\nu^2>$  &  1.74 & 1.61 & 1.59 & & 1.74 & 1.66 &1.58\\
$R_{\mathrm{d}}(R_{\mathrm{L1}})$  & $0.67\pm0.05$ &  $0.73\pm0.04$ & $0.73\pm0.03$ & & $0.71\pm0.04$ & $0.73\pm0.03$ & $0.75\pm0.02$ \\
$\alpha$           & $0.17\pm0.09$ &  $0.45\pm0.13$  & $0.60\pm0.13$  & & $0.19\pm0.10$ & $0.32\pm0.11$ & $0.51\pm0.12$ \\
$T_1$($10^3$~K)    & $14.8\pm2.05$ & $23.0\pm4.44$ & $25.6\pm4.77$ & & $14.7\pm4.52$ & $18.5\pm4.51$ & $24.0\pm5.84$ \\
$T_{\mathrm{d}}$($10^3$~K)    & $7.22\pm0.97$ & $8.76\pm1.91$ & $8.92\pm1.77$ & & $8.15\pm1.68$ & $8.35\pm1.82$ & $10.2\pm2.40$ \\
$\chi_\mathrm{s}$              & $1.18\pm0.15$ & $1.27\pm0.11$ & $1.23\pm0.08$ & & $1.14\pm0.14$ & $1.29\pm0.13$ & $1.32\pm0.11$ \\
$(B-V)_{\circ}$    & $0.35\pm0.15$ & $0.13\pm0.19$ & $0.08\pm0.14$ & & $0.26\pm0.17$ & $0.18\pm0.19$ & $0.03\pm0.16$  \\
\enddata
\end{deluxetable}

\clearpage

\begin{deluxetable}{lccccccc}
\tablenum{9}
\rotate
\tablecaption{Intermediate Polar Distribution Mean Solutions}
\tablewidth{0pt}
\tablehead{
& & \colhead{$q=0.20$} & & \colhead{} & & \colhead{$q=0.40$} & \\ 
\colhead{Parameter} & \colhead{$h_{\mathrm{r}}=0.00$} & \colhead{$h_{\mathrm{r}}=0.04$} & \colhead{$h_{\mathrm{r}}=0.08$} & \colhead{} & \colhead{$h_{\mathrm{r}}=0.00$} & \colhead{$h_{\mathrm{r}}=0.04$} & \colhead{$h_{\mathrm{r}}=0.08$}} 
\startdata
$<\chi_\nu^2>$  &  1.77 & 1.67  &1.69 & &1.73&1.65 &1.62\\
$R_{\mathrm{d}}(R_{\mathrm{L1}})$      & $0.65\pm0.05$ &  $0.68\pm0.05$ & $0.65\pm0.05$ & & $0.69\pm0.05$ & $0.72\pm0.04$ & $0.72\pm0.03$ \\
$\alpha$           & $0.42\pm0.15$ &  $0.60\pm0.14$  & $0.60\pm0.15$  & & $0.35\pm0.15$ & $0.55\pm0.16$ & $0.65\pm0.12$ \\
$T_1$($10^3$~K)    & $15.2\pm2.33$ & $23.8\pm4.11$ & $24.7\pm4.44$ & & $13.2\pm2.57$ & $17.8\pm3.02$ & $24.1\pm4.42$ \\
$T_{\mathrm{d}}$($10^3$~K)    & $6.54\pm0.88$ & $6.66\pm0.94$ & $6.53\pm0.88$ & & $6.82\pm0.98$ & $6.94\pm1.10$ & $7.36\pm1.31$ \\
$\chi_\mathrm{s}$              & $1.18\pm0.15$ & $1.16\pm0.08$ & $1.10\pm0.06$ & & $1.15\pm0.14$ & $1.20\pm0.13$ & $1.18\pm0.10$ \\
$(B-V)_{\circ}$    & $0.30\pm0.10$ & $0.26\pm0.13$ & $0.30\pm0.12$ & & $0.31\pm0.14$ & $0.23\pm0.16$ & $0.18\pm0.14$  \\
\enddata
\end{deluxetable}

\clearpage

\begin{deluxetable}{lccccccc}
\tablenum{10}
\rotate
\tablecaption{Luminosities, Mass Accretion Rates, and Distances}
\tablewidth{0pt}
\tablehead{
& & \colhead{$q=0.20$} & & \colhead{} & & \colhead{$q=0.40$} & \\ 
\colhead{Parameter} & \colhead{$h_{\mathrm{r}}=0.00$} & \colhead{$h_{\mathrm{r}}=0.04$} & \colhead{$h_{\mathrm{r}}=0.08$} & \colhead{} & \colhead{$h_{\mathrm{r}}=0.00$} & \colhead{$h_{\mathrm{r}}=0.04$} & \colhead{$h_{\mathrm{r}}=0.08$}} 
\startdata
$\eta_\mathrm{d,s}$ & 155 (148) & 387 (283) & 3178 (493) & & 98.4 (83.4) & 148 (146) & 622 (277) \\
$\eta_\mathrm{r,d}$ & 0.0 (0.0) & 0.34 (0.34) & 0.51 (0.40) & & 0.0 (0.0) & 0.21 (0.17) & 0.31 (0.19) \\
$M_\mathrm{V}$ & 7.9 (7.9) & 6.9 (7.2) & 4.6 (6.6) & & 8.4 (8.6) & 7.9 (7.9) & 6.4 (7.2) \\
$\dot M_{17}~(\mathrm{g~s}^{-1})$ & 1.49 (2.34) & 2.26 (2.83) & 17.4 (3.38) & & 1.68 (2.37) & 2.05 (2.84) & 6.47 (3.36) \\
$\dot M_\mathrm{crit,17}~(\mathrm{g~s}^{-1})$ & 1.72 (2.34) & 2.50 (1.72) & 2.50 (2.09) & & 1.82 (1.50) & 2.20 (1.82) & 2.20 (2.20) \\
$f_2$  & 0.23 (0.26) & 0.077 (0.11) & 0.0089 (0.056) & & 0.16 (0.19) & 0.088 (0.096) & 0.022 (0.051) \\ 
$(m-M)_\mathrm{V}$ & 8.34 (8.20) & 9.52 (9.14) & 11.87 (9.87) & & 8.73 (8.54) & 9.38 (9.29) & 10.88 (9.97) \\
$A_\mathrm{V}$ & 0.0 (0.12) & 0.45 (0.30) & 1.71 (0.78) & & 0.03 (0.12) & 0.33 (0.45) & 1.23 (0.81) \\
$d~(\mathrm{kpc})$ & 0.47 (0.42) & 0.65 (0.59) & 1.08 (0.66) & & 0.55 (0.48) & 0.65 (0.59) & 0.85 (0.68) \\ 
\enddata
\end{deluxetable}

\clearpage

\begin{figure}
\epsscale{0.80}
\plotone{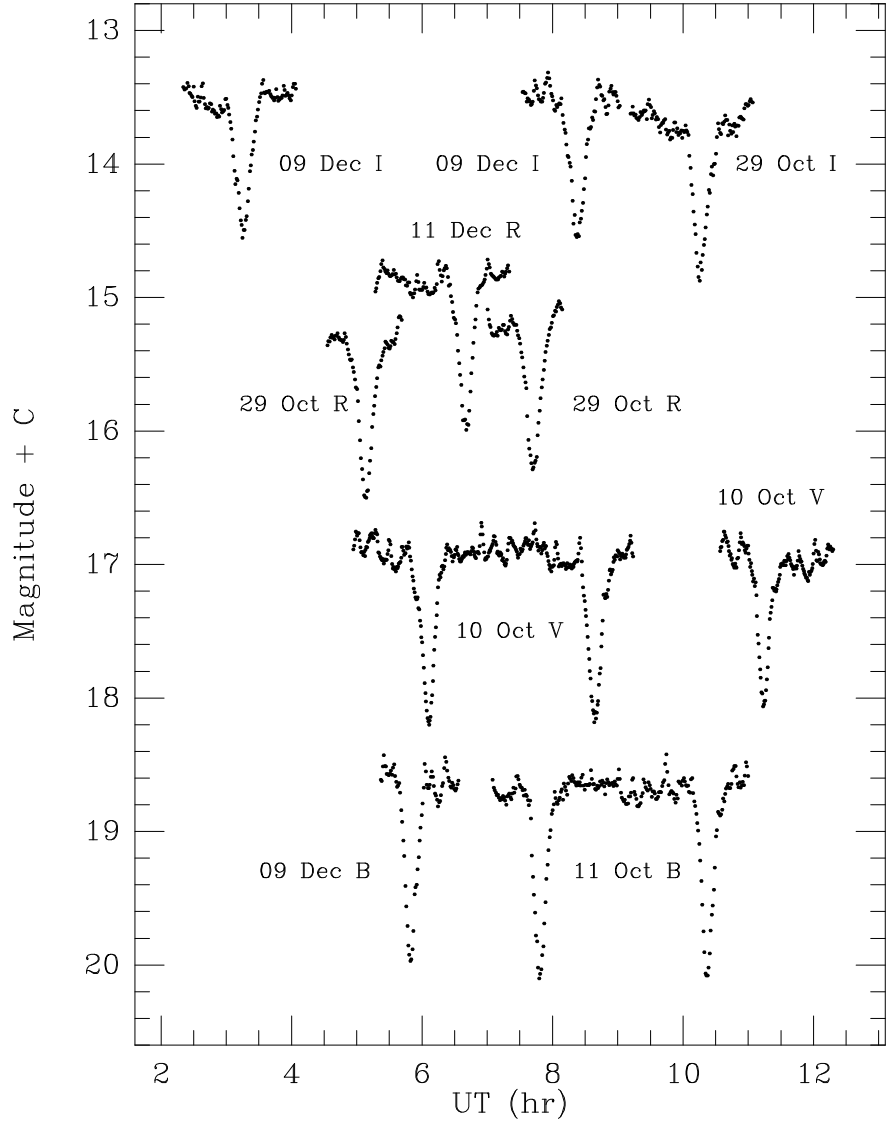}
\caption{The multicolor light curves of V~Per. For clarity of
          presentation, the data have been
          offset by addition of the following constants: $B$-band: C=0.0;
          $V$-band: C=1.2; $R$-band: C=2.4; $I$-band: C=4.0.}
\end{figure}

\begin{figure}
\epsscale{0.80}
\plotone{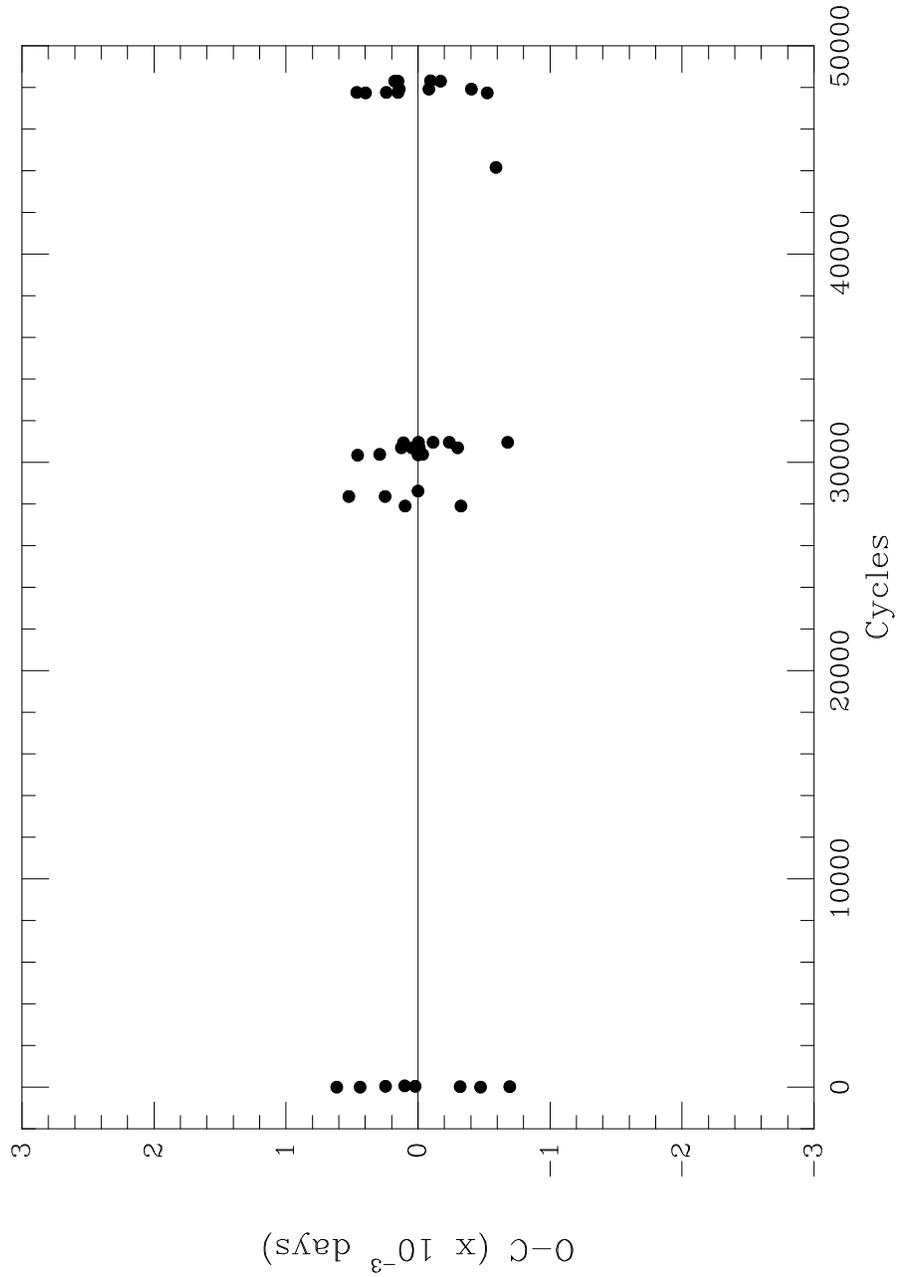}
\caption{The residuals of the the observed times of mid-eclipse
          with respect to eqn~(1) is plotted as a function of
          cycle number. The final grouping of points near cycle number
          48000 are based on eclipse timings from this work, while the
          remaining points are based on timings from the literature.
          There is no evidence for any period change in V~Per over the
          $\sim15$ years of available eclipse timings.}
\end{figure}

\begin{figure}
\epsscale{0.80}
\plotone{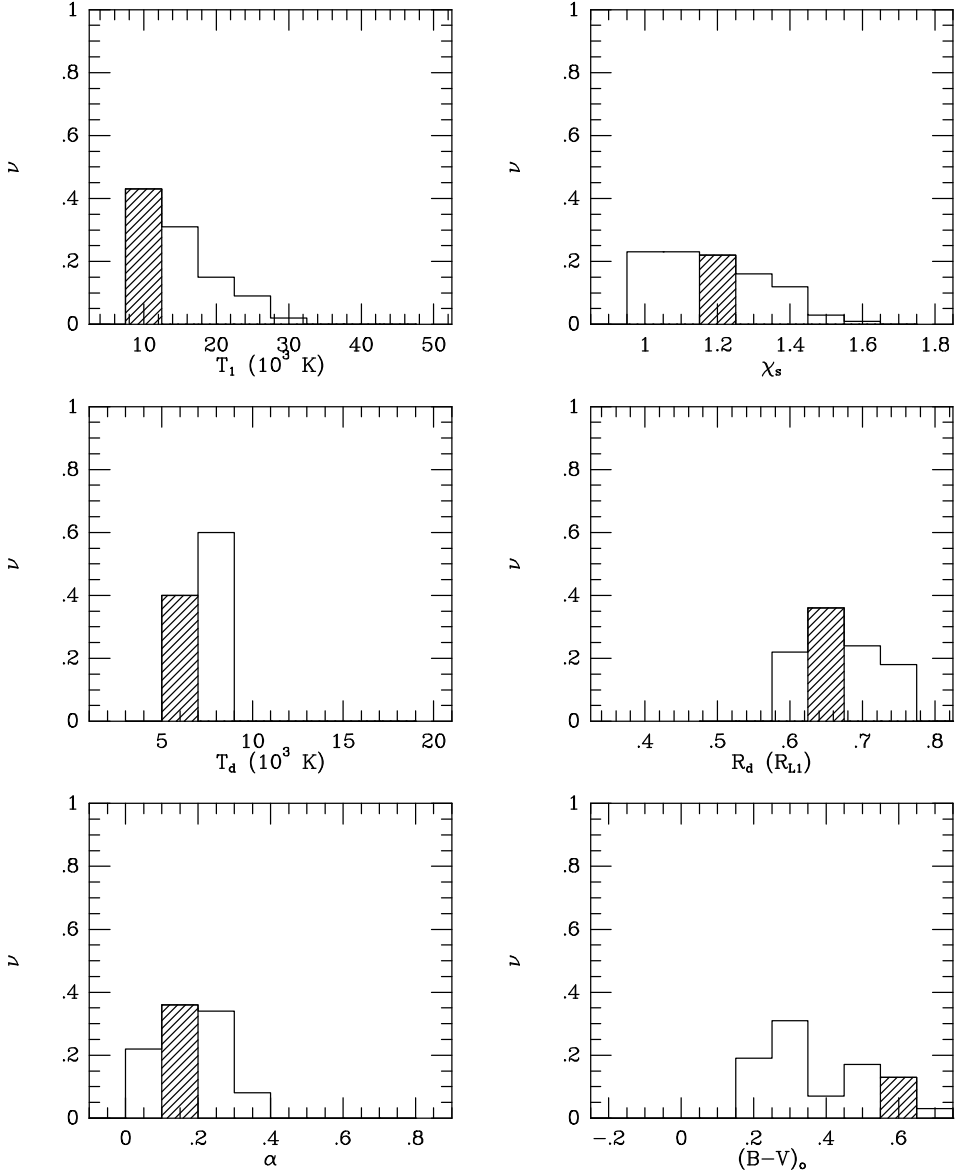}
\caption{The $q=0.20, h_{\mathrm{r}}=0.00$
          frequency distributions for each fitting parameter.
          The range of each parameter has been chosen
          to include all combinations of parameters that result in
          acceptable fits of the top 100 best-fitting models to the data.
          The cross-hatched regions indicate the parameter values
          that produce the optimum fit to the data.}
\end{figure}

\begin{figure}
\epsscale{0.80}
\plotone{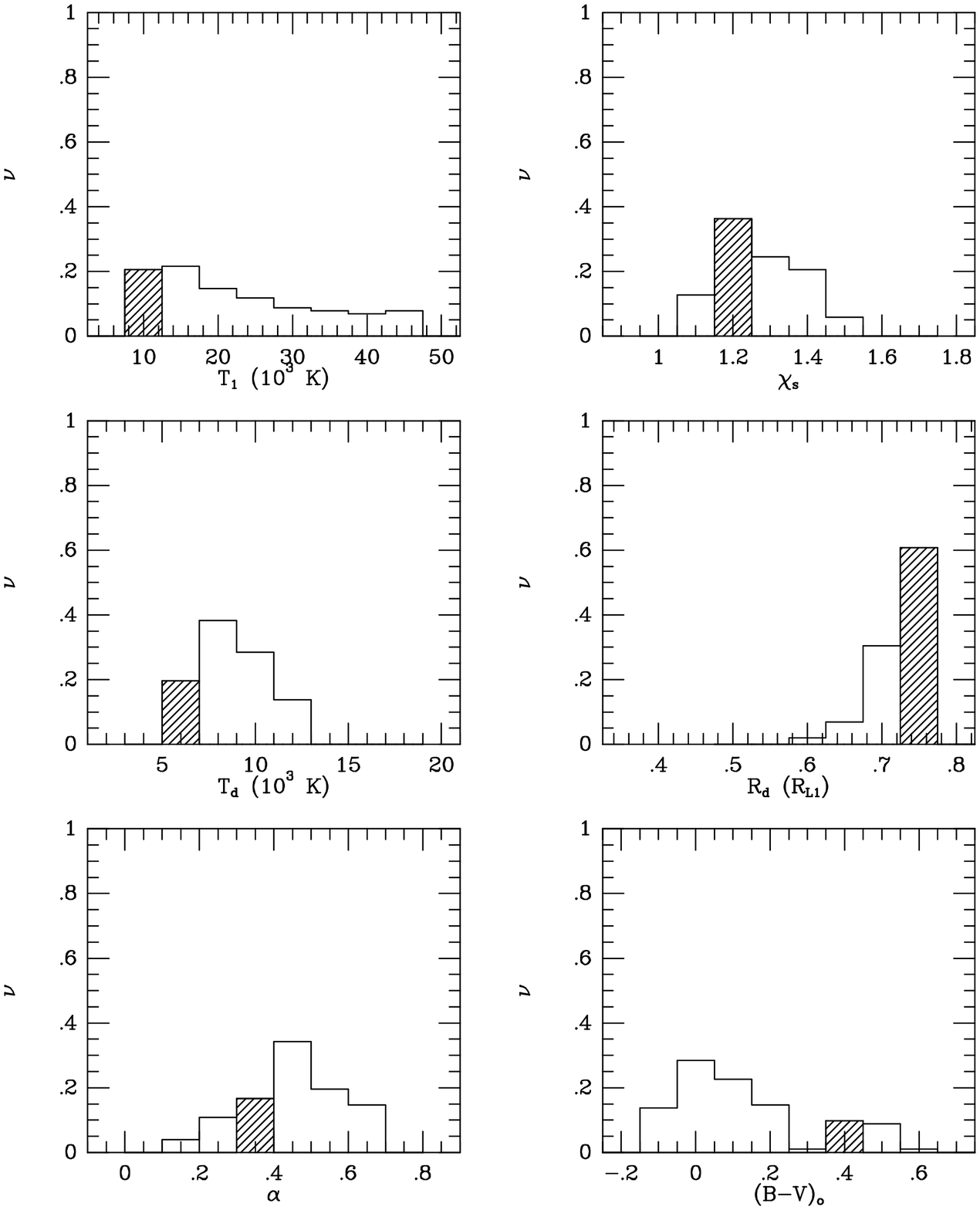}
\caption{The same as Figure~3, but for $h_{\mathrm{r}}=0.04$.}
\end{figure}

\begin{figure}
\epsscale{0.80}
\plotone{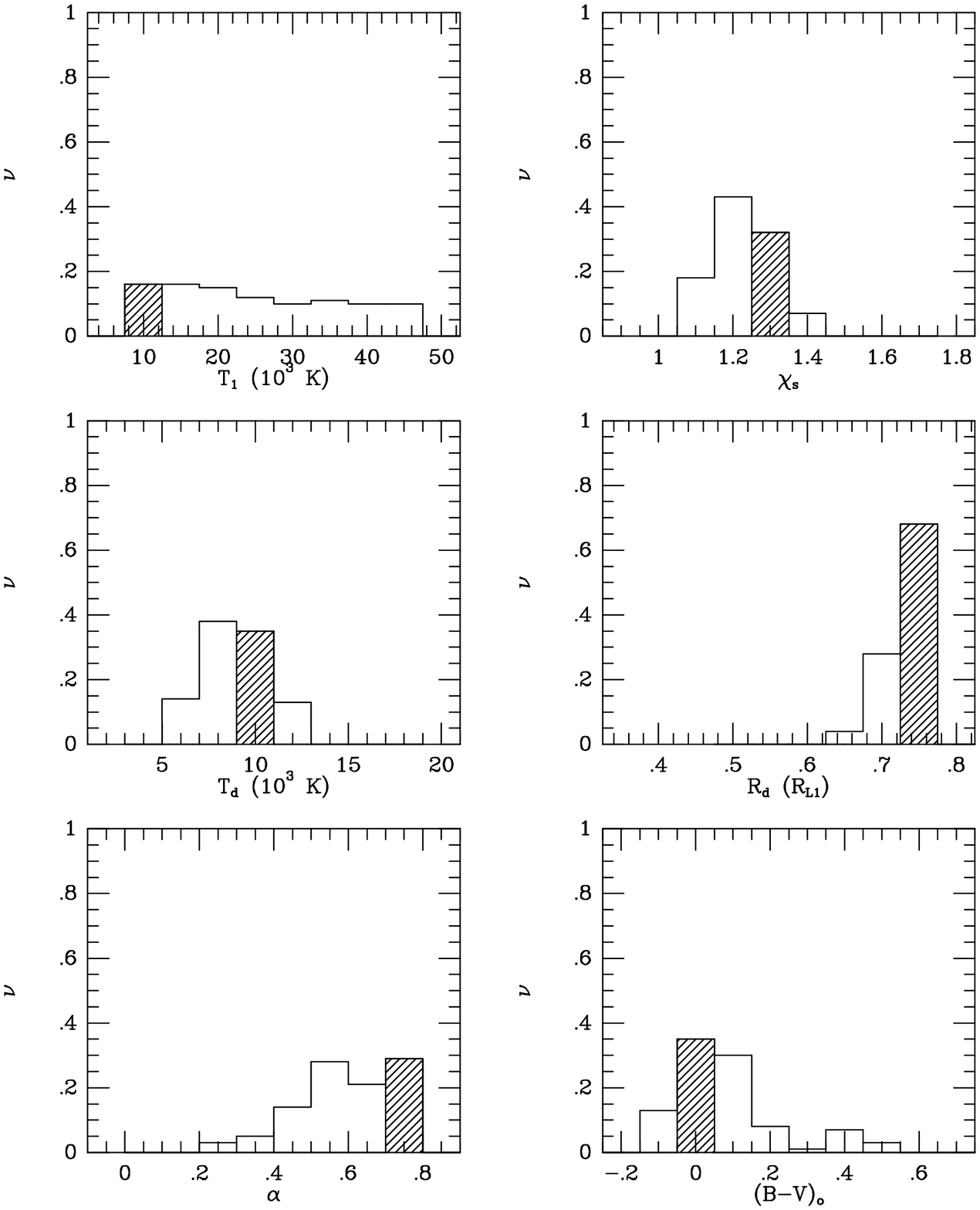}
\caption{The same as Figure~3, but for $h_{\mathrm{r}}=0.08$.}
\end{figure}

\begin{figure}
\epsscale{0.80}
\plotone{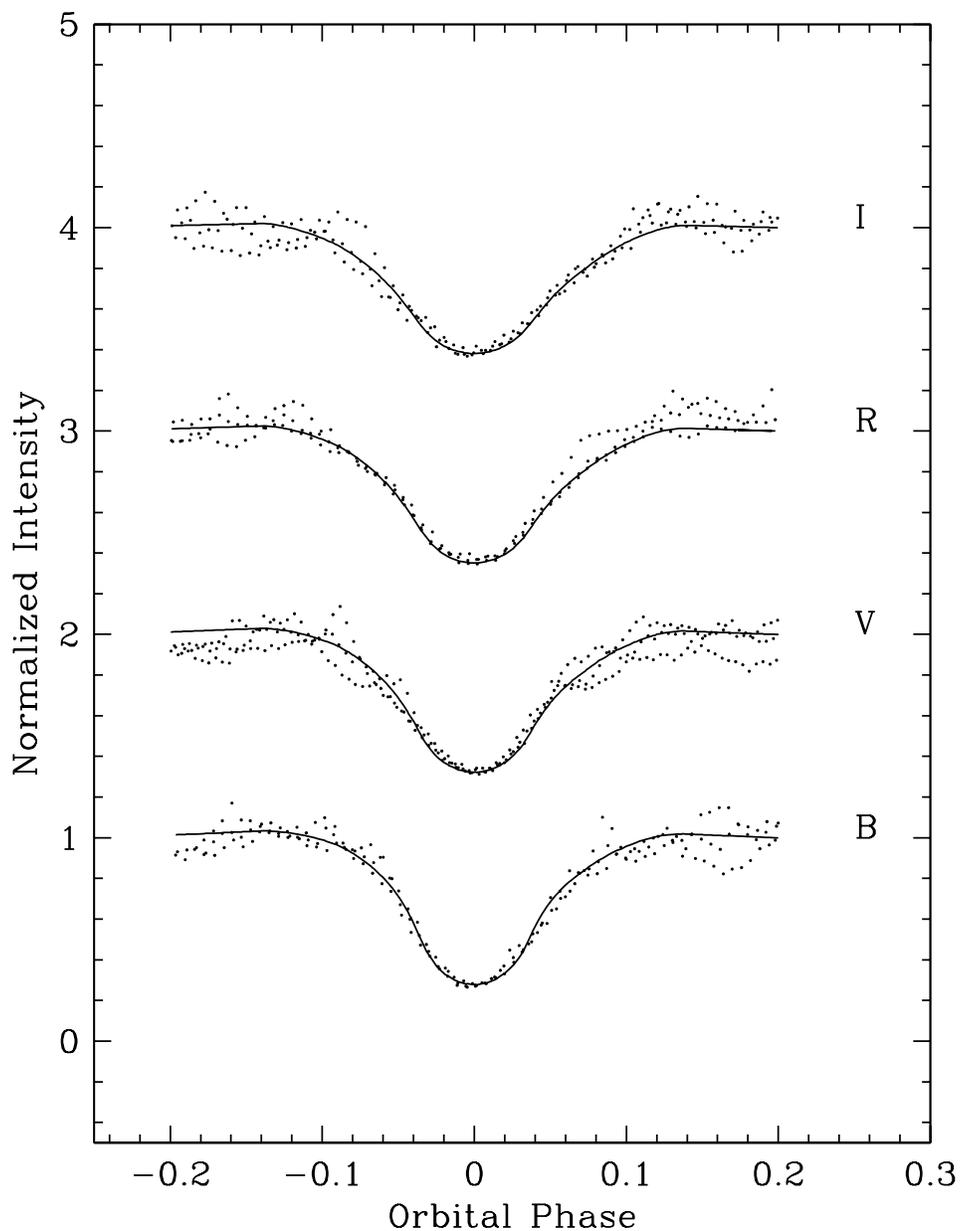}
\caption{The best-fitting eclipse profiles (solid lines) for
          the $q=0.20, h_\mathrm{r}=0.08$ model are plotted
          together with the observed data. The data have been converted to
          relative intensity and normalized to unity in the orbital phase
          range: $0.15 < \phi < 0.20$.
          For clarity of presentation the $V$, $R$, and $I$-band
          light curves and models have been shifted
          upward by constant offsets of 1, 2, and 3, respectively.}
\end{figure}

\begin{figure}
\epsscale{0.80}
\plotone{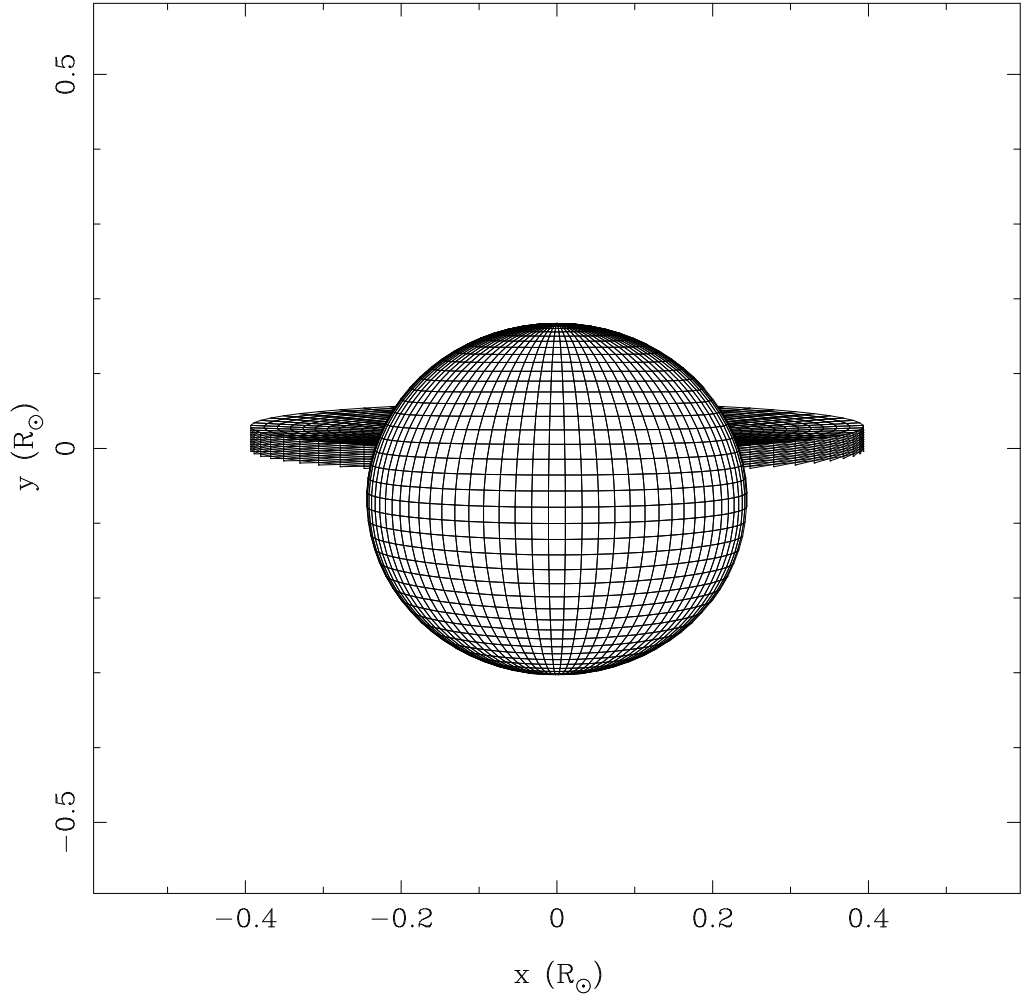}
\caption{A schematic representation of the V~Per geometry at mid-eclipse
          for the $q=0.2, i=85.4, h_{\mathrm{r}}=0.04$ model. Note that the
          accretion disk is never completely eclipsed, and that the outer
          disk and rim
          make a significant contribution to the light at mid-eclipse.}
\end{figure}

\begin{figure}
\epsscale{0.80}
\plotone{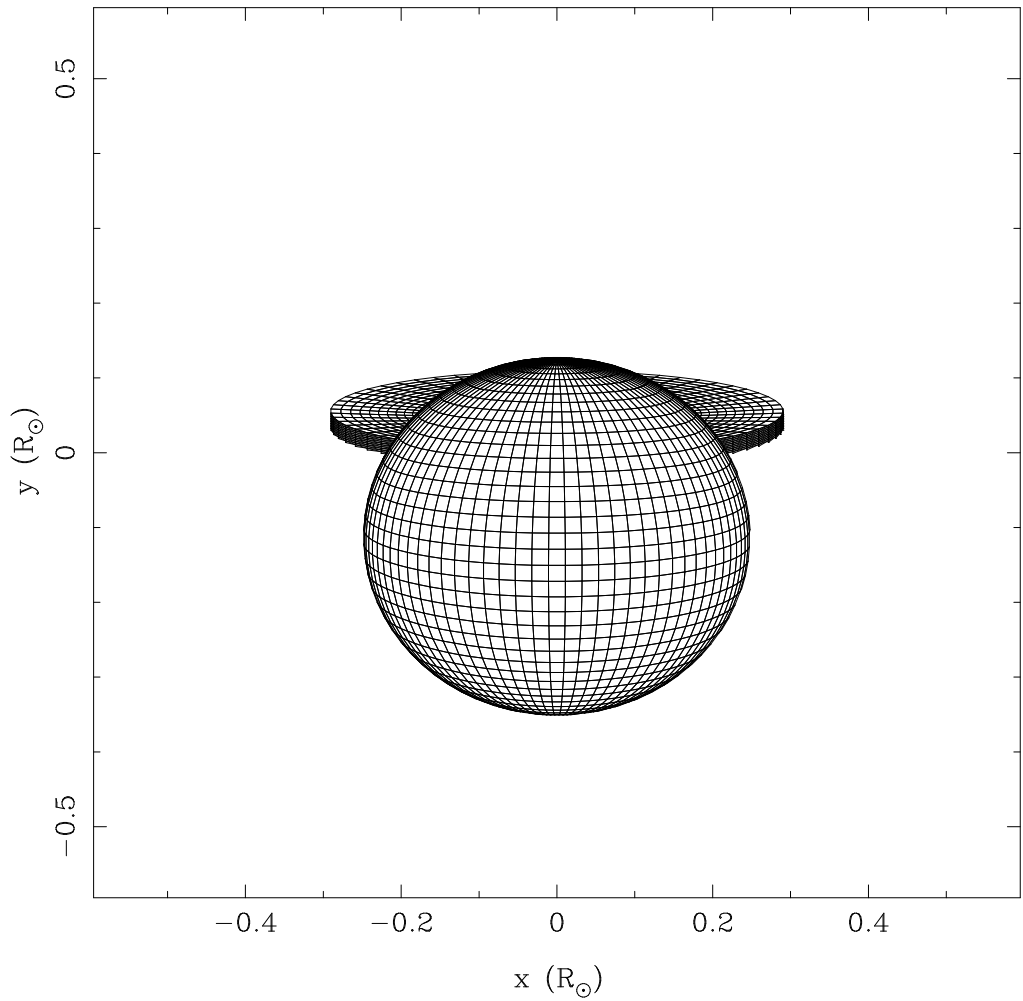}
\caption{The same as Figure~7, but for $q=0.4, i=79.4, h_{\mathrm{r}}=0.04$.}
\end{figure}

\begin{figure}
\epsscale{0.80}
\plotone{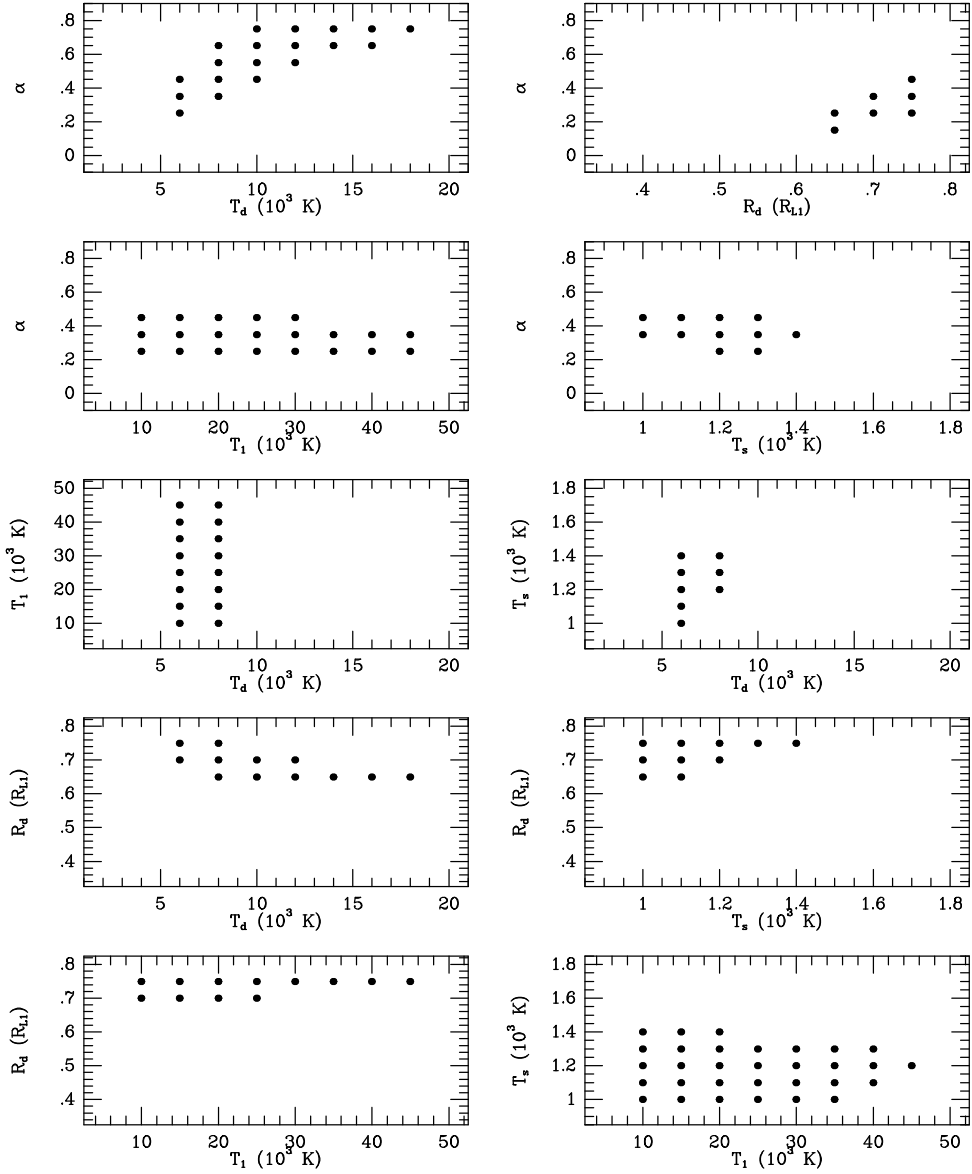}
\caption{The correlations between pairs of model parameters for the
          representative $q=0.20, h_{\mathrm{r}}=0.04$ model
          are shown
          for each of the 10 possible pairings of the five fitting
          parameters. In each case a range of model solutions ($\chi_\nu^2 < 2.0$)
          for each pair of parameters is plotted, while the remaining
          three parameters are held fixed at their optimum values (Table~7).}
\end{figure}

\end{document}